# Power Control with Imperfect Exchanges and Applications to Spectrum Sharing

Nikolaos Gatsis and Georgios B. Giannakis, *Fellow, IEEE*

**Abstract**

In various applications, the effect of errors in gradient-based iterations is of particular importance when seeking saddle points of the Lagrangian function associated with constrained convex optimization problems. Of particular interest here are problems arising in power control applications, where network utility is maximized subject to minimum signal-to-interference-plus-noise ratio (SINR) constraints, maximum interference constraints, maximum received power constraints, or simultaneous minimum and maximum SINR constraints. Especially when the gradient iterations are executed in a disributed fashion, imperfect exchanges among the link nodes may result in erroneous gradient vectors. In order to assess and cope with such errors, two running averages (ergodic sequences) are formed from the iterates generated by the perturbed saddle point method, each with complementary strengths. Under the assumptions of problem convexity and error boundedness, bounds on the constraint violation and the suboptimality per iteration index are derived. The two types of running averages are tested on a spectrum sharing problem with minimum and maximum SINR constraints, as well as maximum interferece constraints.

## I. Introduction

The gradient projection method has well-documented merits for finding saddle points of the Lagrangian function associated with general constrained convex optimization problems, and specifically those arising in power control applications. Of particular importance in this context is the effect of errors (perturbations) in the gradient vectors on the performance of the method.

In these applications, a set of links comprising transmitter-receiver pairs communicate over the same frequency band, causing interference to each other. The resulting signal-to-interference-plus-noise ratio

Manuscript received August 23, 2010. Work in this paper was supported by NSF grants CCF-0830480 and ECCS-0824007. Parts have been submitted to the *IEEE Int. Conf. Acoustics, Speech, and Signal Processing*, Prague, Czech Republic, 2011.

N. Gatsis and G. B. Giannakis are with the Department of Electrical and Computer Engineering, University of Minnesota, USA (e-mails: {gatsisn, georgios}@umn.edu).

(SINR) is the main indicator of each link's quality-of-service (QoS). The aim is to regulate the transmit-powers so that a sum of utility functions of the SINR across links is maximized, under constraints, which may include minimum QoS support in terms of SINR, maximum interference constraints, maximum receive-power constraints, or, a combination of application-dependent minimum *and* maximum QoS expressed in terms of SINR; see e.g., [1]–[5].

The constrained nature of the problem naturally leads to solution methods based on the Lagrangian function. A dual method cannot be readily employed, because the coupling of power variables renders minimization of the Lagrangian function at each time slot difficult. Instead, gradients of the Lagrangian function with respect to the primal variables and Lagrange multipliers are used in order to find a saddle point of the Lagrangian; see e.g., [6, Sec. 3.5], [7], [8] and also [1], [4] for applications to power control. In order for these gradients to be made available at the various nodes, some exchange of information is necessary through the reversed network or message passing—see e.g., [9, Ch. 6], [1] for the former, and [10], [5] for the latter. As exchanges may be imperfect, they result in erroneous gradients.

Errors in optimization methods have been the object of considerable research. Specifically, errors in the gradient method for finding saddle points have been adressed in the context of stochastic approximation [11], [12], where a strictly convex objective, diminishing stepsizes, or linearly independent active constraint gradients are typically assumed. Primal methods with erroneous gradient or subgradient vectors have also received attention. Bounded deterministic errors are considered in [13], [14]. Random errors are studied in [15]; and in [16], using running averages (ergodic sequences). Various deterministic or random error models are considered in [17, Ch. 4 and Sec. 5.5].

In the context of power control, network utility maximization in the absence of power-coupling constraints has been pursued in [18] using the stochastic approximation framework, and diminishing stepsizes. Unconstrained optimization with logarithmic utilities and errors in the knowledge of link gains has been studied in [10]. Even in the absense of errors, ergodic sequences are formed in order to obtain constrained Nash equilbria in power control games [19].

This paper deals with the assessment and mitigation of errors in the gradient method for saddle points, in contrast to [13]–[17], which deal with primal methods. Relative to stochasic approximation methods [11], [12], the errors here are modeled differently—namely as bounded but otherwise arbitrary—and explicit per iteration bounds are developed on the induced constraint violation and suboptimality. Due to their general setting, the results are valuable within the theme of error analysis in optimization theory.

The specific contributions and organization of the paper are as follows. Section II presents two collections of power control problems: (a) typical problems where network utility is maximized subject to minimum QoS constraints, maximum interference, or, receive-power constraints; and (b) contemporary spectrum sharing problems. In the latter, primary users (license holders) and secondary users simultaneously access a licensed band, or all users are allowed to use an unlicensed band [20]. Simultaneous minimum and maximum QoS constraints in terms of SINR must be imposed in both cases; maximum interference constraints are also incorporated here, extending related results of [4].

Section III begins with description of the gradient method for saddle points in order to obtain a unifying solver for the various power control formulations. Message passing or the reversed network approach are employed to distribute the gradient iterations. Then, focus turns to error-resilient gradient-based iterations and their analysis. The impact of errors is mitigated and analyzed through running averages of the iterates obtained by the perturbed saddle point method. Two types of averages are considered, namely: (a) one with equal weights for all iterates; and (b) one for which past iterates are weighted via exponentially decaying weights. The analysis is applicable when constant stepsize is used—which is desirable in resource allocation algorithms—and under persistent (non-vanishing) errors. Each type of averaging has its own merits. Explicit bounds are derived per iteration on the constraint violation and the suboptimality induced by errors. Related results but for the first type of averaging and *without* errors can be found in [8].

Finally, numerical tests are presented in Section IV, and conclusions with pointers to future directions in Section V.

## II. UNIFYING POWER CONTROL FORMULATIONS

### A. Typical Power Control Problems

Consider the power control problem for a single-channel (i.e., single-carrier) network in which users share the same frequency band, e.g., as in CDMA. Assuming a peer-to-peer operating setup, there is a set of $\mathcal{M} := \{1, \ldots, M\}$ links, where each link $i \in \mathcal{M}$ comprises a dedicated transmitter ($\text{Tx}_i$) wishing to communicate with a corresponding receiver ($\text{Rx}_i$). The terms pair, user, and link will be used interchangeably. Let $h_{ij}$ denote the (power) path gain from $\text{Tx}_i$ to $\text{Rx}_j$, that is assumed invariant. The path gain $h_{ij}$ models the relationship between the transmitted and received power, and captures any signal processing operation taking place at the transmitter or the receiver. Also, let $n_i$ denote the noise power at $\text{Rx}_i$; $p_i$ the transmit-power of $\text{Tx}_i$; and $p_i^{\max}$ the maximum power budget $\text{Tx}_i$ can afford, i.e., $0 \leq p_i \leq p_i^{\max}$. The received SINR $\gamma_i$ at $\text{Rx}_i$ is a function of the powers $\boldsymbol{p} := [p_1, \ldots, p_M]^T$ given by

$$\gamma_i := \frac{h_{ii} p_i}{n_i + \sum_{k \neq i} h_{ki} p_k}. \tag{1}$$

Model (1) is general, and can accommodate a number of (de-) modulation schemes [9, Ch. 4]. For future use, define vectors $\boldsymbol{p}^{\max} := [p_1^{\max}, \ldots, p_M^{\max}]^T$, $\boldsymbol{\gamma} := [\gamma_1, \ldots, \gamma_M]^T$; and the matrix $\mathbf{A} = [a_{ij}]$ with $a_{ij} := h_{ji}/h_{ii}$ if $i \neq j$, and $a_{ii} := 0$.

The utility associated with each link $i \in \mathcal{M}$ will be described by a generic function $u_i(\gamma_i)$. The goal is to maximize the sum of all link utilities subject to QoS, interference, or receive-power constraints. For all optimization problems described next, the following two operating conditions are adopted.

**C1.** Utilities $u_i(\gamma_i)$, $i = 1, \ldots, M$, are chosen so that: (a) they are strictly increasing and twice continuously differentiable; and (b) $-\gamma_i u_i''(\gamma_i)/u_i'(\gamma_i) \geq 1$ for $\gamma_i > 0$ ($'$ denotes differentiation).

**C2.** The noise power satisfies $n_i > 0$ for all $i$; and the gain matrix $\mathbf{A}$ is irreducible; see e.g., [9, Def. A.27].

Condition C1 is standard in the power control literature to guarantee that $u_i(\gamma_i)$ is concave in $\ln \gamma_i$; see e.g., [9, Ch. 5], [21]. It also effects the fairness condition $\lim_{\gamma_i \to 0^+} u_i(\gamma_i) = -\infty$, which guarantees that *non-zero* power is allocated to all users. Examples of utilities satisfying C1 are $u_i(\gamma_i) = \ln \gamma_i$, $u_i(\gamma_i) = \gamma_i^\alpha/\alpha$ with $\alpha < 0$, and $u_i(\gamma_i) = \ln[\ln(1 + \gamma_i)]$. Furthermore, the irreducibility of $\mathbf{A}$ in C2 is also a standard assumption in power control problems [1]. It means the users cannot be divided in two disjoint groups without at least one user in one group being interfered by a user in the other group.

Power control must also account for the following constraints; see e.g., [1]–[3].

1) *Minimum QoS support:* To ensure minimum QoS levels, the QoS per link $i$ is generically described here by a function $v_i(\gamma_i)$, which can e.g., represent rate when $v_i(\gamma_i) = \ln(1 + \gamma_i)$. If $v_i(\gamma_i)$ is chosen monotonic, then minimum constraints on $v_i$ map one-to-one to SINR bounds, $\gamma_i \geq \gamma_i^{\min}$.
2) *Maximum interference constraints:* The interference-plus-noise (IpN) term $n_i + \sum_{k \neq i} h_{ki} p_k$ at $\mathrm{Rx}_i$ [cf. (1)] is constrained not to exceed $q_i^{\max}$. In this way for instance, weak links are protected from excessive interference.
3) *Maximum receive-power constraints:* The total received power $h_{ii} p_i + n_i + \sum_{k \neq i} h_{ki} p_k$ at $\mathrm{Rx}_i$ is constrained not to exceed $s_i^{\max}$. Wireles mess networks is an application, whereby different wireless systems sharing the same bandwidth may each connect to a different node in the mesh network, and regulating the received power helps accomodating more such systems.

Not all previously mentioned types of constraints are necessarily present simultaneously. However,

retaining them all leads to the following unifying form of the power control problem:

$$\max_{\mathbf{0}\leq\boldsymbol{p}\leq\boldsymbol{p}^{\max}} \quad \sum_{i=1}^{M} u_i(\gamma_i) \tag{2a}$$

$$\text{subj. to} \quad \gamma_i^{\min} \leq \gamma_i \quad \forall i \in \mathcal{M} \tag{2b}$$

$$n_i + \sum_{k\neq i} h_{ki}p_k \leq q_i^{\max} \quad \forall i \in \mathcal{M} \tag{2c}$$

$$h_{ii}p_i + n_i + \sum_{k\neq i} h_{ki}p_k \leq s_i^{\max} \quad \forall i \in \mathcal{M}. \tag{2d}$$

If the trasformation $p_i = e^{y_i}$ is applied, then (2) becomes a convex optimization problem in $\boldsymbol{y} := [y_1, \ldots, y_M]^T$. Specifically, the objective is a concave function of $\boldsymbol{y}$; see e.g., [9, Ch. 6]. The power constraints become $\boldsymbol{y} \in \mathcal{Y}$, where $\mathcal{Y} := \{\boldsymbol{y} \in \mathbb{R}^M | y_i \leq \ln p_i^{\max},\ i = 1, \ldots, M\}$. Constraints (2b)–(2d) take the form of sums of exponentials, which are convex; see e.g., [22, Ch. 3].

### B. The Spectrum Sharing Paradigm

The spectrum sharing paradigm has been put forward by the Federal Communications Commission, and allows a frequency to be utilized by more users than the ones licensed [20]. In this paradigm, there are two models of interest here.

- *Flexible primary model:* License holders, called primary users (PUs), allow secondary users (SUs) to access the spectrum. The SUs pay a fee to the PUs, and receive QoS in a range determined by this fee. This is oftentimes called a secondary market. The PUs should also be guaranteed a minimum QoS or maximum interference level.
- *Open sharing model:* All users are considered primary, and cooperate in order to achieve efficient resource management. To this end, they voluntarily set lower and upper bounds on the received QoS.

It is apparent that in both models, the resource allocation task must account for minimum and maximum bounds on the QoS, and possibly maximum interference constraints. Formulations are presented next for single-channel and multi-channel settings.

*1) Single-Channel Networks:* Recalling the notation of Subsection II-A, upper and lower bounds on QoS expressed in terms of $v_i(\gamma_i)$ map one-to-one to SINR bounds; i.e., $v_i(\gamma_i) \in [v_i(\gamma_i^{\min}), v_i(\gamma_i^{\max})] \Leftrightarrow \gamma_i \in [\gamma_i^{\min}, \gamma_i^{\max}]$. Moreover, an upper bound $q_i^{\max}$ on the interference $n_i + \sum_{k\neq i} h_{ki}p_k$ inflicted to link

$i$ may be considered. Putting things together, the associated power control problem amounts to

$$\max_{\mathbf{0}\le\boldsymbol{p}\le\boldsymbol{p}^{\max}} \quad \sum_{i=1}^{M} u_i(\gamma_i) \tag{3a}$$

$$\text{subj. to} \quad \gamma_i^{\min} \le \gamma_i \le \gamma_i^{\max} \quad \forall i \in \mathcal{M} \tag{3b}$$

$$n_i + \sum_{k\neq i} h_{ki} p_k \le q_i^{\max} \quad \forall i \in \mathcal{M}. \tag{3c}$$

It is now worth checking how the general formulation (3) can be useful in describing the design objectives in the spectrum sharing models previously described.

**Application 1** (Flexible primary with minimum QoS guarantees)**.** The set of users is divided into a set of PUs $\mathcal{M}_p$, and a set of SUs $\mathcal{M}_s$, i.e., $\mathcal{M} = \mathcal{M}_p \cup \mathcal{M}_s$. The PUs set bounds on the received QoS of the SUs, based on the fee that the latter pay. Moreover, minimum SINR guarantees are also included for the PUs. Thus, constraint (3b) is specialized to $\gamma_i^{\min} \le \gamma_i$ for all $i \in \mathcal{M}_p$, and $\gamma_i^{\min} \le \gamma_i \le \gamma_i^{\max}$ for all $i \in \mathcal{M}_s$. Constraint (3c) is absent. The SUs may or may not have minimum SINR guarantees, depending on the agreement between them and the PUs.

**Application 2** (Flexible primary with interference protection)**.** The PU-SU paradigm is adopted as in Application 1; but now, the PUs are protected against excessive interference. Constraint (3c) takes the form $n_i + \sum_{k\neq i} h_{ki} p_k \le q_i^{\max}$ for all $i \in \mathcal{M}_p$, while (3b) becomes $\gamma_i^{\min} \le \gamma_i \le \gamma_i^{\max}$ for all $i \in \mathcal{M}_s$.

**Application 3** (Open sharing model)**.** In the open sharing model, all users are peers, and cooperate in order to achieve efficient allocation of the network resources. Specifically, each user can set bounds matching its own requirements on Qos, leading to (3b). Constraint (3c) is absent.

Note that the transformation $p_i = e^{y_i}$—which has been the "workhorse" for efficient power control, as seen in the previous subsection—does not readily convexify (3). The reason is that the second constraint in (3b) becomes the superlevel set of the sum-exp function. In order to facilitate the solution of (3) through convex optimization, the following operating condition is adopted, on top of C1 and C2.

**C3.** If every user has a maximum SINR constraint, there is no power vector $\tilde{\boldsymbol{p}}$ with $\mathbf{0} < \tilde{\boldsymbol{p}} \le \boldsymbol{p}^{\max}$ and $n_i + \sum_{k\neq i} h_{ki} \tilde{p}_k \le q_i^{\max}$ for all $i \in \mathcal{M}$ such that the resulting SINRs $\tilde{\gamma}_i$ satisfy $\tilde{\gamma}_i = \gamma_i^{\max}$ for all $i \in \mathcal{M}$.

The condition is satisfied automatically for flexible primary models, because primary users in this case do not set upper bounds on the received QoS. For the open sharing model, the condition can be easily checked using the standard power control algorithm of [23].

Having clarified the operating conditions, we proceed to relax the nonconvex (3). To this end, let $q_i$ denote an auxiliary variable associated with link $i$, upper-bounding the interference-plus-noise (IpN) term $n_i + \sum_{k\neq i} h_{ki}p_k$. Collecting all variables $q_i$ in $\boldsymbol{q} := [q_1, \ldots, q_M]^T$ and constraints $q_i^{\max}$ in $\boldsymbol{q}^{\max} := [q_1^{\max}, \ldots, q_M^{\max}]^T$, consider the following relaxed version of (3):

$$\max_{\mathbf{0}\leq\boldsymbol{p}\leq\boldsymbol{p}^{\max},\, \boldsymbol{q}\leq\boldsymbol{q}^{\max}} \quad \sum_{i=1}^{M} u_i(h_{ii}p_i q_i^{-1}) \tag{4a}$$

$$\text{subj. to} \quad \gamma_i^{\min} \leq h_{ii}p_i q_i^{-1} \leq \gamma_i^{\max} \quad \forall\, i \in \mathcal{M} \tag{4b}$$

$$q_i \geq n_i + \sum_{k\neq i} h_{ki}p_k \qquad \forall\, i \in \mathcal{M}. \tag{4c}$$

It is worth noting that the upper bound $q_i^{\max}$ is now placed on the variable $q_i$.

Problem (4) can be transformed into an equivalent convex optimization problem. To this end, apply the one-to-one change of variables $p_i = e^{y_i}$ and $q_i = e^{z_i}$, and define $\boldsymbol{y} := [y_1, \ldots, y_M]^T$ and $\boldsymbol{z} := [z_1, \ldots, z_M]^T$. Under C1, problem (4) is equivalent to a convex optimization problem in $(\boldsymbol{y}, \boldsymbol{z})$. The convex set constraint for $\boldsymbol{y}$ is $\mathcal{Y}$ as in the formulations of Subsection II-A, while for $\boldsymbol{z}$, it is $\mathcal{Z} = \{\boldsymbol{z} \in \mathbb{R}^M | z_i \leq \ln q_i^{\max}\, \forall i \in \mathcal{M}\}$.

The ensuing lemma proved in Appendix A asserts that the optimal solution of (4) satisfies (4c) with equality, and therefore, it is the solution of (3).

**Lemma 1.** *If* (3) *is feasible and C1–C3 hold, the optimal solution* $\boldsymbol{p}^*$ *,*$\boldsymbol{q}^*$ *of* (4)*, satisfies*

$$q_i^* = n_i + \sum_{k\neq i} h_{ki}p_k^* \quad \forall i \in \mathcal{M}. \tag{5}$$

The convexity and optimality of the relaxed problem (4) will be leveraged in Section III. The previous ideas are generalized to multi-channel networks in what follows.

*2) Multi-Channel Networks:* Users here may transmit over an orthogonal set of frequency bands $\mathcal{F} := \{1, \ldots, F\}$, also referred to as channels, subcarriers, or tones. The power of $\text{Tx}_i$ on channel $f$ is $p_{i,f}$, the noise power at $\text{Rx}_i$ on channel $f$ is $n_{i,f}$, and the (power) path gain from $\text{Tx}_i$ to $\text{Rx}_j$ on channel $f$ is $h_{ij,f}$. Moreover, each user adheres to a *spectral mask* $p_{i,f} \leq p_{i,f}^{\max}$, and maximum power budget $\sum_f p_{i,f} \leq p_i^{\max}$. Vector $\boldsymbol{p}_i := [p_{i,1}, \ldots, p_{i,F}]^T$ contains the power loadings for user $i$. Then, each user's power must lie in

$$\mathcal{P}_i := \left\{ \boldsymbol{p}_i \geq \mathbf{0} \,\middle|\, p_{i,f} \leq p_{i,f}^{\max} \;\; \forall f \in \mathcal{F};\;\; \sum_{f\in\mathcal{F}} p_{i,f} \leq p_i^{\max} \right\}. \tag{6}$$

The received SINR at $\text{Rx}_i$ on channel $f$ is $\gamma_{i,f} := h_{ii,f}p_{i,f}/(n_{i,f} + \sum_{k\neq i} h_{ki,f}p_{k,f})$. Similar to the single-channel case, $\mathbf{A}_f$ is the gain matrix for channel $f$.

The aim is to formulate the power control problem for a multi-channel network incorporating diverse QoS constraints as well as interference constraints. To this end, utility functions of the SINR per user and channel are adopted, namely, $u_{i,f}(\gamma_{i,f})$, $U_{i,f}(\gamma_{i,f})$, and $V_{i,f}(\gamma_{i,f})$. QoS is collected for each user by summing each type of utility function across channels. The first one is used to give the objective function to be maximized, the second one to impose a lower bound $U_i^{\min}$, and the third one to set an upper bound $V_i^{\max}$. Regarding the interference constraints, there are two ways to generalize (3c): imposing either (a) individual bounds $q_i^{\max}$ per user across channels, or, (b) individual bounds $q_{i,f}^{\max}$ per user and channel. The choice between the two possibly depends on the bandwidth of each channel; case (a) may be more suitable for smaller channel bandwidths, and (b) for larger channel bandwidths.

All in all, the optimization problem generalizing (3) to multi-channel networks is

$$\max_{\boldsymbol{p}_i \in \mathcal{P}_i\, \forall i \in \mathcal{M}} \sum_{i=1}^{M} \sum_{f=1}^{F} u_{i,f}(\gamma_{i,f}) \tag{7a}$$

$$\text{subj. to} \sum_{f=1}^{F} U_{i,f}(\gamma_{i,f}) \geq U_i^{\min} \text{ and } \sum_{f=1}^{F} V_{i,f}(\gamma_{i,f}) \leq V_i^{\max},\ i \in \mathcal{M} \tag{7b}$$

$$\sum_{f=1}^{F} \left( n_{i,f} + \sum_{k \neq i} h_{ki,f} p_{k,f} \right) \leq q_i^{\max} \text{ or } n_{i,f} + \sum_{k \neq i} h_{ki,f} p_{k,f} \leq q_{i,f}^{\max}, f \in \mathcal{F}, i \in \mathcal{M}. \tag{7c}$$

Clearly, (7) can be specialized appropriately to obtain multi-channel counterparts of Applications 1–3.

Similar to the single-channel case, a relaxation of (7) is formulated. Let $\boldsymbol{q}_i := [q_{i,1}, \ldots, q_{i,F}]^T$ be the local IpN vector. Moreover, define the following constraint sets for $\boldsymbol{q}_i$, corresponding to the two versions of (7c):

$$\mathcal{Q}_i := \left\{ \boldsymbol{q}_i \geq \boldsymbol{0} \,\middle|\, \sum_{f \in \mathcal{F}} q_{i,f} \leq q_i^{\max} \right\} \text{ or } \mathcal{Q}_i := \left\{ \boldsymbol{q}_i \geq \boldsymbol{0} \,\middle|\, q_{i,f} \leq q_{i,f}^{\max}\ \forall f \in \mathcal{F} \right\}. \tag{8}$$

To form the relaxation of (7), the SINR $\gamma_{i,f}$ is replaced by the ratio $h_{ii,f} p_{i,f} / q_{i,f}$, and the local IpN constraint $q_{i,f} \geq n_{i,f} + \sum_{j \neq i} h_{ji,f} p_{j,f}$ is introduced. The transformation is now $p_{i,f} = e^{y_{i,f}}$ and $q_{i,f} = e^{z_{i,f}}$ for all $i$ and $f$, and define $\boldsymbol{y}_i := [y_{i,1}, \ldots, y_{i,F}]^T$, $\boldsymbol{z}_i := [z_{i,1}, \ldots, z_{i,F}]^T$, and $\boldsymbol{y}$ and $\boldsymbol{z}$ as the vectors collecting $\boldsymbol{y}_i$ and $\boldsymbol{z}_i$ for all $i \in \mathcal{M}$, respectively. The transformed constraint set $\mathcal{Y}_i$ for the powers $\boldsymbol{y}_i$ comprises the inequalities $y_{i,f} \leq \ln p_{i,f}^{\max}$ and $\sum_{f \in \mathcal{F}} e^{y_{i,f}} \leq p_i^{\max}$ [cf. (6)], while $\mathcal{Y} := \mathcal{Y}_1 \times \ldots \times \mathcal{Y}_M$. The transformed constraint set $\mathcal{Z}_i$ for the variables $\boldsymbol{z}_i$ takes the form of $\sum_{f=1}^{F} e^{z_{i,f}} \leq q_i^{\max}$ or $z_{i,f} \leq \ln q_{i,f}^{\max}$ [cf. (8)]; define also $\mathcal{Z} := \mathcal{Z}_1 \times \ldots \times \mathcal{Z}_M$.

The following operating conditions, corresponding to C1–C3, are adopted.

**C1′.** Utilities $u_{i,f}(\gamma_{i,f})$ and $U_{i,f}(\gamma_{i,f})$, $i = 1, \ldots, M$, satisfy C1. Utilities $V_{i,f}(\gamma_{i,f})$, $i = 1, \ldots, M$, are chosen so that: (a) they are strictly increasing and twice continuously differentiable; and (b) they are concave and satisfy $-\gamma_{i,f} V''_{i,f}(\gamma_{i,f})/V'_{i,f}(\gamma_{i,f}) \leq 1$ for $\gamma_{i,f} > 0$.

**C2′.** It holds that $n_{i,f} > 0$ for all $i$ and $f$, and gain matrix $\mathbf{A}_f$ is irreducible for all $f$.

**C3′.** If every user has a maximum utility constraint, there are no $\tilde{\boldsymbol{p}}_i$, $\tilde{\boldsymbol{q}}_i$ with $\tilde{\boldsymbol{p}}_i \in \mathcal{P}_i$, $\tilde{\boldsymbol{q}}_i \in \mathcal{Q}_i$ such that the maximum QoS constraint holds with equality for all $i$ [cf. (7b) with $\gamma_{i,f}$ replaced by $h_{ii,f} p_{i,f}/q_{i,f}$].

It is easy to verify that the transformed relaxed problem is a convex optimization problem under C1′. Moreover, the relaxation incurs no loss of optimality under C1′–C3′, analogously to Lemma 1.

## III. Unifying Algorithm and Error Analysis

All power control problems described in the previous section share the characteristic that they include QoS, interference, or received power constraints, which couple the power variables. A dual method for finding the solution is not readily applicable, because this coupling renders maximizing the Lagrangian difficult. For this reason, the optimal power allocation is solved by a gradient method, which yields a saddle point of the associated Lagrangian function.

In order to facilitate the development, the power control problems of Section II can be put in the following generic form.[1]

$$\min_{\boldsymbol{x} \in \mathcal{X}} \quad f(\boldsymbol{x}) \tag{9a}$$

$$\text{subj. to} \quad \boldsymbol{g}(\boldsymbol{x}) \leq \mathbf{0}. \tag{9b}$$

The optimization variable $\boldsymbol{x}$ is $\boldsymbol{y}$ for problem (2); and $\mathcal{X} = \mathcal{Y}$. In the case of (3) or (7), $\boldsymbol{x}$ collects the respective $\boldsymbol{y}$ and $\boldsymbol{z}$, and similarly, $\mathcal{X}$ is $\mathcal{Y} \times \mathcal{Z}$. The association of functions $f(\boldsymbol{x}) : \mathbb{R}^N \to \mathbb{R}$ and $\boldsymbol{g}(\boldsymbol{x}) : \mathbb{R}^N \to \mathbb{R}^K$ with the objective and the constraint functions in the power control problems with appropriate choice of dimensions $N$ and $K$, is evident.

Let $\boldsymbol{\zeta}$ denote a vector of Lagrange multipliers corresponding to constraints (9b). The Lagrangian function of problem (9) is

$$L(\boldsymbol{x}, \boldsymbol{\zeta}) = f(\boldsymbol{x}) + \boldsymbol{\zeta}^T \boldsymbol{g}(\boldsymbol{x}). \tag{10}$$

The following assumption is adopted for problem (9).

[1] All references to (2) will in fact be to its convex equivalent after the transformation $p_i = e^{y_i}$. All references to (3) and (7) will be to their relaxed convex versions with variables $(\boldsymbol{y}, \boldsymbol{z})$. Moreover, it will be clear when the symbol $f$ will be used to denote a channel in the multi-channel power control problem, or, the objective function in the generic optimization problem.

**Assumption 1.** Functions $f(\boldsymbol{x})$ and $\boldsymbol{g}(\boldsymbol{x})$ are convex and continuous on $\mathcal{X}$, and the set $\mathcal{X}$ is convex, closed, and bounded. The set of optimal solutions $\mathcal{X}^*$ is nonempty, closed and bounded. Moreover, Slater constraint qualification holds, i.e., there is an $\check{\boldsymbol{x}} \in \mathcal{X}$ so that $\boldsymbol{g}(\check{\boldsymbol{x}}) < \mathbf{0}$.

Assumption 1 is clearly satisfied by the power control problems of Section II. In particular, convexity is ensured by condition C1. The optimal powers are non-zero due to C1, and remain constrained by maximum power budgets, hence entries of the optimal $\boldsymbol{y}^*$ are bounded. In fact, straightforward application of Weierstrass' theorem [24, Prop. 2.1.1] implies that the optimal sets are closed and bounded in all power control problems of Section II. Slater's constraint qualification was not explicitly mentioned in the previous section, but it is a natural assumption for the power control problems. The following observation is an immediate consequence of Assumption 1; see e.g., [24, Ch. 6] for the related theory. ($\|.\|$ denotes the Euclidean norm, and $\mathbb{R}_+$ the nonnegative reals).

*Observation.* The set of optimal primal solutions $\mathcal{X}^*$ and the set of optimal dual solutions (optimal Lagrange multipliers) $\mathcal{D}^*$ are convex, as solution sets of convex optimization problems. Set $\mathcal{D}^*$ is also nonempty, closed, and bounded. In particular, there is constant $B^*_{\boldsymbol{\zeta}} > 0$ so that $\|\boldsymbol{\zeta}^*\| \leq B^*_{\boldsymbol{\zeta}}$ for any dual optimal $\boldsymbol{\zeta}^*$. Furthermore, for any convex set $\mathcal{D}$ with $\mathcal{D}^* \subset \mathcal{D} \subset \mathbb{R}^K_+$, the pairs $(\boldsymbol{x}^*, \boldsymbol{\zeta}^*)$ with any $\boldsymbol{x}^* \in \mathcal{X}^*$ and $\boldsymbol{\zeta}^* \in \mathcal{D}^*$ are exactly the saddle points of the Lagrangian function (10) over $\mathcal{X} \times \mathcal{D}$.

Based on this observation, solving (9) amounts to finding a saddle point of the associated Lagrangian function over $\mathcal{X} \times \mathcal{D}$. Let $\mathsf{P}_{\mathcal{X}}$ and $\mathsf{P}_{\mathcal{D}}$ denote the projection on $\mathcal{X}$ and $\mathcal{D}$, respectively; and $\nabla_{\boldsymbol{x}} L(\boldsymbol{x}, \boldsymbol{\zeta})$ and $\nabla_{\boldsymbol{\zeta}} L(\boldsymbol{x}, \boldsymbol{\zeta})$ the gradient of the Lagrangian with respect to $\boldsymbol{x}$ and $\boldsymbol{\zeta}$, respectively. Consider the following gradient projection algorithm, indexed by $t = 0, 1, 2, \ldots$, with $\alpha > 0$ a constant stepsize:

$$\boldsymbol{x}(t+1) = \mathsf{P}_{\mathcal{X}}[\boldsymbol{x}(t) - \alpha \nabla_{\boldsymbol{x}} L(\boldsymbol{x}(t), \boldsymbol{\zeta}(t))] \tag{11a}$$

$$\boldsymbol{\zeta}(t+1) = \mathsf{P}_{\mathcal{D}}[\boldsymbol{\zeta}(t) + \alpha \nabla_{\boldsymbol{\zeta}} L(\boldsymbol{x}(t), \boldsymbol{\zeta}(t))]. \tag{11b}$$

A desirable feature of the algorithm (11) is the *constant stepsize*, which enables adaptation in resource allocation algorithms where the parameters may vary slowly. Another feature is the projection, which here ensures that the transmit powers remain *within budget* in each and every iteration.

In order to perform the updates (11) in a distributed fashion suitable for power control, the partial derivatives of the Lagrangian must become available at the nodes they are needed through exchange of information, which may entail errors. The ensuing Subsection III-A outlines methods of exchanging information among nodes, and asserts the convergence of (11) in the error-free case. The latter is relevant if sufficiently strong error control codes are used, to ensure error-free exchanges. Then, Subsection III-B pursues error-resilient iterations and their performance when exchanges are imperfect. Subsection III-C explains how to select certain performance-critical parameters involved in the context of power control. The material in Subsection III-A unifies ideas from [1] and [4], and extends them also to include the more general formulations of Subsection II-B. The results of Subsection III-B are of interest also outside the context of power control, as they are developed for a general optimization algorithm.

### A. Error-free exchanges

Iterations (11) do not converge in general, even under convexity assumptions. In the context of power control, the objective (2a) is strictly convex under C1 and C2 [1]. For problem (4), strict convexity of the objective function does not hold in general. (For instance, the utility $u_i(\gamma_i) = \ln \gamma_i$ under the transformation $p_i = e^{y_i}$, $q_i = e^{z_i}$ becomes $\sum_{i=1}^{M}(\ln h_{ii} + y_i - z_i)$.) However, the following weaker property is shown to hold in Appendix A.

**Lemma 2.** *Under C1–C3 (or C1′–C3′ for the multi-channel problems) and Slater's constraint qualification, the following property holds for the Lagrangian functions of* (3) *and* (7) *for any* $\boldsymbol{\zeta}^* \in \mathcal{D}^*$*:*

$$L(\boldsymbol{x}^*, \boldsymbol{\zeta}^*) < L(\boldsymbol{x}, \boldsymbol{\zeta}^*) \quad \forall \boldsymbol{x} \in \mathcal{X},\, \boldsymbol{x} \neq \boldsymbol{x}^*. \tag{12}$$

This property is called stability of the saddle points with respect to $\boldsymbol{x}$ in [7]. Clearly, (12) holds automatically if $f(\boldsymbol{x})$ is strictly convex.

Either strict convexity of $f(\boldsymbol{x})$ or (12) alone ensure convergence of (11) by immediate application of [7]; see also [1] and [4] for related results in power control. With $\Omega^*$ denoting the set of optimal primal solutions-Lagrange multipliers $\mathcal{X}^* \times \mathcal{D}^*$, and $\operatorname{dist}(\boldsymbol{\omega}, \Omega)$ the distance of a point $\boldsymbol{\omega} := (\boldsymbol{x}, \boldsymbol{\zeta})$ from a closed convex set $\Omega$, the convergence claim is summarized next.[2]

**Lemma 3.** *Suppose C1–C3 (or C1′–C3′ for the multi-channel problems) and Slater's constraint qualification hold, implying Assumption 1 and* (12)*. For every* $\varepsilon$ *and* $\delta$ *with* $0 < \varepsilon < \delta$*, there exist positive* $\alpha_0(\varepsilon, \delta)$ *and* $t_0(\varepsilon, \delta)$ *such that for any stepsize* $0 < \alpha \leq \alpha_0(\varepsilon, \delta)$*, and any initial point* $\boldsymbol{\omega}(0) \in \mathcal{X} \times \mathcal{D}$ *with* $\operatorname{dist}(\boldsymbol{\omega}(0), \Omega^*) \leq \delta$*, the iterates* $\boldsymbol{\omega}(t)$ *in* (11) *satisfy* $\operatorname{dist}(\boldsymbol{\omega}(t), \Omega^*) \leq \varepsilon$ *for all* $t \geq t_0(\varepsilon, \delta)/\alpha$*. This further implies that* $\operatorname{dist}(\boldsymbol{x}(t), \mathcal{X}^*) \leq \varepsilon$ *for all* $t \geq t_0(\varepsilon, \delta)/\alpha$*.*

This lemma asserts that the power allocation will remain as close as desired to the optimal one after a number of iterations using a sufficiently small stepsize—the stepsize and the number of iterations depend

[2]The distance from a closed convex set is defined as $\operatorname{dist}(\boldsymbol{\omega}, \Omega) := \inf_{\boldsymbol{w} \in \Omega} \|\boldsymbol{\omega} - \boldsymbol{w}\|$ [24, p. 88].

on the desired proximity of $\boldsymbol{x}(t)$ to $\mathcal{X}^*$.

Next, the distributed implementation of (11) is considered. The updates of $y_i$, $z_i$, and the associated Lagrange multipliers pertaining to link $i$ are performed at Tx$_i$. For this to be possible, Tx$_i$ needs to know the partial derivatives of the Lagrangian function with respect to $y_i$ and $z_i$. It is easy to verify that all these derivatives, except $\partial L(\boldsymbol{\omega})/\partial y_i$, depend on the channel $h_{ii}$ and the current SINR value, namely, $\gamma_i(t) = h_{ii}e^{y_i(t)}/(n_i + \sum_{j\neq i} h_{ji}e^{y_j(t)})$. The latter can be fed back from Rx$_i$—where it is measured—to the corresponding Tx$_i$.

On the other hand, $\partial L(\boldsymbol{\omega})/\partial y_i$ for each of the formulations of Section II involves a sum that depends on quantities non-local to link $i$. This term has the form $e^{y_i(t)} \sum_{j\neq i} h_{ij}\xi_j(t)$, where $\xi_j$ takes the following values.

1) For the case (2), let $\mu_j$, $\lambda_j$, and $\nu_j$, $j = 1, \ldots, M$, be Lagrange multipliers corresponding to the convex form of constraints (2b)–(2d), respectively. Then,

$$\xi_j(t) = \underbrace{u'_j(\gamma_j(t))\frac{\gamma_j^2(t)}{h_{jj}e^{y_j(t)}} + \frac{\mu_j(t)}{h_{jj}e^{y_j(t)}}}_{A} + \underbrace{u'_j(\gamma_j(t))\frac{\gamma_j^2(t)}{h_{jj}e^{y_j(t)}} + \lambda_j(t)}_{B} + \underbrace{u'_j(\gamma_j(t))\frac{\gamma_j^2(t)}{h_{jj}e^{y_j(t)}} + \nu_j(t)}_{C} \tag{13}$$

where the terms $A$, $B$, $C$ correspond to the parts of the Lagrangian for constraints (2b)–(2d), respectively.

2) For the case (3),

$$\xi_j(t) = \mu_j(t)e^{-z_j(t)} \tag{14}$$

where $\mu_j$, $j = 1, \ldots, M$, are the Lagrange multipliers corresponding to the convex form of constraints (4c), that is, $e^{-z_j}(n_j + \sum_{k\neq j} h_{kj}e^{y_k}) - 1 \leq 0$. The multi-channel case (7) is analogous, where now the non-local part has the form $\sum_f e^{y_i^f} \sum_{j\neq i} h_{ij}^f \mu_j^f e^{-z_j^f}$, and $\mu_j^f$, $j = 1, \ldots, M$, $f = 1, \ldots, F$, correspond to the convex form of the local IpN constraints.

Note that $\xi_j(t)$ depends on quantities which are known to Tx$_j$, $j \neq i$. There are two ways to make $\sum_{j\neq i} h_{ij}\xi_j(t)$ available to Tx$_i$: (a) message passing, and (b) the reversed network operation.

In the *message passing* approach, Tx$_i$ acquires the cross-channels $h_{ij}$ to all other receivers through training and/or feedback. At each time slot, Tx$_j$ must then broadcast the variable $\xi_j(t)$ to all other transmitters.

In the *reversed network* approach, no exchange of information among different links is required, but the links are assumed reciprocal; that is, the channel from Tx$_i$ to Rx$_j$ is identical to the channel from Rx$_j$ to Tx$_i$. The transmitters become receivers, and vice-versa. In order to use the reversed network, rewrite $\sum_{j \neq i} h_{ij}\xi_j(t)$ as $\sum_{j=1}^{M} h_{ij}\xi_j(t) - h_{ii}\xi_i(t)$. The sum $\sum_{j=1}^{M} h_{ij}\xi_j(t)$ is then the received power at $\text{Tx}_i$ (which is a receiver for the reversed network), if all receivers (which are transmitters for the reversed network) transmit symbols with power $\xi_j(t)$. In order for the receivers of the links to know the quantities $\xi_j(t)$, they need to perform updates of all local quantities, starting from the same initialization as their corresponding transmitter.

For the multi-channel case, the previously described operations are performed on a per-channel basis. In any of the two distributed implementations, the value of $\partial L(\boldsymbol{\omega})/\partial y_i$ [or $\partial L(\boldsymbol{\omega})/\partial y_i^f$] may contain errors, induced by perturbed exchanges among links. This motivates the study of (11) in a more general setting in the ensuing subsection, where the gradients of the Lagrangian are perturbed by errors.

### B. Imperfect Exchanges

Suppose $\nabla_{\boldsymbol{x}} L(\boldsymbol{x}(t), \boldsymbol{\zeta}(t))$ is perturbed by error $\boldsymbol{r}(t)$ at iteration $t$, while $\nabla_{\boldsymbol{\zeta}} L(\boldsymbol{x}(t), \boldsymbol{\zeta}(t))$ is perturbed by error $\boldsymbol{\epsilon}(t)$. Then, iterations (11) become

$$\boldsymbol{x}(t+1) = \mathsf{P}_{\mathcal{X}}[\boldsymbol{x}(t) - \alpha(\nabla_{\boldsymbol{x}} L(\boldsymbol{x}(t), \boldsymbol{\zeta}(t)) + \boldsymbol{r}(t))] \tag{15a}$$

$$\boldsymbol{\zeta}(t+1) = \mathsf{P}_{\mathcal{D}}[\boldsymbol{\zeta}(t) + \alpha(\nabla_{\boldsymbol{\zeta}} L(\boldsymbol{x}(t), \boldsymbol{\zeta}(t)) + \boldsymbol{\epsilon}(t))]. \tag{15b}$$

Iterations (15) and their performance pursued afterwards have the following desirable attributes:

a1) Persistent (i.e., nonvanishing) errors are accounted for.

a2) Constant stepsize is used, which enables adaptability.

a3) The error-induced constraint violation and suboptimality can be bounded per iteration.

Key to establishing a1)–a3) is usage of suitable averages of the iterates $\{\boldsymbol{x}(t)\}$. These can help cope with nonvanishing errors and constant stepsizes, both of which drive the terms $\alpha r(t)$ and $\alpha\epsilon(t)$ in (15) to stay away from zero. Consider the following running average [8]

$$\bar{\boldsymbol{x}}(t) := \frac{1}{t}\sum_{i=0}^{t-1} \boldsymbol{x}(i), \quad t = 1, 2, \ldots \tag{16}$$

and also its exponentially weighted counterpart with the so-called forgetting factor $\beta \in (0, 1)$

$$\bar{\boldsymbol{x}}_\beta(t) := \frac{\sum_{i=0}^{t-1} \beta^{t-1-i}\boldsymbol{x}(i)}{\sum_{i=0}^{t-1} \beta^{t-1-i}} = \frac{\sum_{i=0}^{t-1} \beta^{-i}\boldsymbol{x}(i)}{\sum_{i=0}^{t-1} \beta^{-i}}, \quad t = 1, 2, \ldots \tag{17}$$

inspired by the exponentially decaying window approaches used in adaptive signal processing [25]. Clearly, averaging as in (17) weighs more the recent iterates. Eq. (16) may be viewed as obtained from (17) in the limit $\beta \to 1$. The sequences $\{\bar{\boldsymbol{x}}(t)\}$ and $\{\bar{\boldsymbol{x}}_\beta(t)\}$ which are formed from the iterates $\{\boldsymbol{x}(t)\}$ are often referred to as ergodic sequences. Being convex combinations of the iterates $\{\boldsymbol{x}(0), \boldsymbol{x}(1), \ldots, \boldsymbol{x}(t-1)\}$, the running averages $\bar{\boldsymbol{x}}(t)$ and $\bar{\boldsymbol{x}}_\beta(t)$ belong to the set $\mathcal{X}$ for all $t \geq 1$. It is important to remark that both running averages can be efficiently computed in a recursive fashion, e.g., $\bar{\boldsymbol{x}}(t)$ can be computed using $\bar{\boldsymbol{x}}(t-1)$ and $\boldsymbol{x}(t)$; and similarly for $\bar{\boldsymbol{x}}_\beta(t)$.

The main result is that the sequences $\{\bar{\boldsymbol{x}}(t)\}$ and $\{\bar{\boldsymbol{x}}_\beta(t)\}$ converge to some neighborhood of the optimal solution of (9). It is further of interest to estimate these neighborhoods—and specifically assess the constraint violation and suboptimality—and also to study how the constraint violation and the objective value evolve across iterations. It will be seen that the two averaging schemes have complementary merits. Specifically, $\{\bar{\boldsymbol{x}}(t)\}$ in (16) may converge to a smaller neighborhood than (17). On the other hand, $\{\bar{\boldsymbol{x}}_\beta(t)\}$ in (17) reduces the constraint violation much faster than (16), which is desirable in practice.

In order to proceed with the development, the ensuing assumptions are adopted in addition to Assumption 1.

**Assumption 2.** The iterates $\{\boldsymbol{x}(t)\}_{t=0}^{\infty}$ and $\{\boldsymbol{\zeta}(t)\}_{t=0}^{\infty}$ generated by (15) are bounded. Specifically, there exists a constant $B_{\boldsymbol{\zeta}} > B_{\boldsymbol{\zeta}}^*$ so that

$$\|\boldsymbol{\zeta}(t)\| \leq B_{\boldsymbol{\zeta}}, \quad t = 0, 1, 2, \ldots \tag{18}$$

Under (18), the gradient vectors are also bounded, i.e., there is a constant $B_L > 0$ so that

$$\|\nabla_{\boldsymbol{x}} L(\boldsymbol{x}(t), \boldsymbol{\zeta}(t))\| \leq B_L,\ \|\nabla_{\boldsymbol{\zeta}} L(\boldsymbol{x}(t), \boldsymbol{\zeta}(t))\| \leq B_L,\ t = 0, 1, 2, \ldots \tag{19}$$

**Assumption 3.** The error sequences are bounded, i.e., there are constants $r > 0$ and $\epsilon > 0$ so that

$$\|\boldsymbol{r}(t)\| \leq r, \quad \|\boldsymbol{\epsilon}(t)\| \leq \epsilon, \quad t = 0, 1, 2, \ldots \tag{20}$$

Assumption 2 will automatically hold if the sets $\mathcal{X}$ and $\mathcal{D}$ are compact, due to the projection on those [cf. (15)]. In power control problems, these sets can be selected to be compact, because the sets of optimal primal variables and optimal Lagrange multipliers are bounded, as explained earlier. In particular, for a constant $\varrho > 0$, the following set can be used as a compact $\mathcal{D}$ for the projection:

$$\mathcal{D}_{\text{ball}} := \{\boldsymbol{\zeta} \geq \mathbf{0} |\ \|\boldsymbol{\zeta}\| \leq B_{\boldsymbol{\zeta}}^* + \varrho\} \tag{21}$$

where $B_{\boldsymbol{\zeta}}^*$ is introduced in Assumption 1. It will be useful to express $B_{\boldsymbol{\zeta}}^*$ using the vector $\check{\boldsymbol{x}}$ satisfying Slater's condition; see also [8]. In particular, let $\check{d}$ denote a lower bound on the optimal dual value of (9), and $g_k(\boldsymbol{x})$, $k = 1, \ldots, K$ the entries of $\boldsymbol{g}(\boldsymbol{x})$. Then, a value for $B_{\boldsymbol{\zeta}}^*$ is

$$B_{\boldsymbol{\zeta}}^* = \frac{f(\check{\boldsymbol{x}}) - \check{d}}{\min_{1 \leq j \leq K}\{-g_j(\check{\boldsymbol{x}})\}}. \tag{22}$$

For the general analysis presented here, it is not necessary to assume that the sets $\mathcal{X}$ and $\mathcal{D}$ are compact. In particular, the analysis holds if $\mathcal{D}$ is chosen as any convex set satisfying $\mathcal{D}^* \subset \mathcal{D} \subset \mathbb{R}_+^K$, or, if $\mathcal{D} = \mathcal{D}_{\text{ball}}$ in particular. In the former case, we can consider under Assumption 2 that the iterates $\{\boldsymbol{\zeta}(t)\}$ will remain in the bounded set given by $\mathcal{D}_{\text{ball}}$ with $\varrho = B_{\boldsymbol{\zeta}} - B_{\boldsymbol{\zeta}}^*$ (with a slight abuse of the notation $\varrho$). It is also worth stressing that when $\mathcal{X}$ and $\mathcal{D}$ are not compact, or when stronger assumptions on the convexity-concavity of the Lagrangian function do not hold, it is natural to assume bounded iterates or gradients; see e.g., [8], [13], [14].

Assumptions 1 and 2 imply that there is a constant $B_d$ so that

$$\|\boldsymbol{x}(t) - \boldsymbol{x}^*\| \leq B_d, \quad \forall t \geq 0,\ \boldsymbol{x}^* \in \mathcal{X}^*. \tag{23}$$

Assumption 3 does not hold when e.g., the errors are randomly drawn from a distribution with infinite support. Such a case may be better handled in the context of stochastic approximations [11], [12]. Note though that strict convexity of the objective function or diminishing stepsizes or linearly independent gradients of active constraints are typically needed for convergence in such a framework. Without invoking such assumptions, Assumption 3 will allow characterization of the constraint violation, and suboptimality assessment as a function of the iteration index. Furthermore, note that Assumption 3 holds if the errors are random deviates from a distribution with bounded support (e.g., uniform) and have arbitrary correlation across time (iteration number), or, across the error vector entries. In this case, the results established here effectively hold with probability 1.

Assumptions 1–3 hold throughout, unless otherwise stated. Let $f^*$ denote the optimal value of (9), and $\boldsymbol{x}^* \in \mathcal{X}^*$ an arbitrary primal optimal solution. The first result provides bounds on the norm of the constraint violation $\|[\boldsymbol{g}(\bar{\boldsymbol{x}}(t))]^+\|$, and the objective value $f(\bar{\boldsymbol{x}}(t))$ at each iteration index $t$ using initialization variables and the quantities defined in (18)–(23) ($[.]^+$ denotes projection onto $\mathbb{R}_+^K$). All proofs of the results in this subsection can be found in Appendix B.

**Proposition 1.** *Under Assumptions 1–3, the sequence $\{\bar{\boldsymbol{x}}(t)\}$ satisfies for $t \geq 1$*

$$\text{(i)}\ \|[\boldsymbol{g}(\bar{\boldsymbol{x}}(t))]^+\| \leq \frac{\|\boldsymbol{\zeta}(0)\|^2 + 2\|\boldsymbol{\zeta}(0)\|B_{\boldsymbol{\zeta}} + B_{\boldsymbol{\zeta}}^2}{2t\alpha\varrho} + \frac{\|\boldsymbol{x}(0) - \boldsymbol{x}^*\|}{2t\alpha\varrho} + \frac{2B_{\boldsymbol{\zeta}}\epsilon + B_d r}{\varrho} + \frac{\alpha(B_L+\epsilon)^2}{2\varrho} + \frac{\alpha(B_L+r)^2}{2\varrho} \tag{24}$$

$$\text{(ii)}\ f(\bar{\boldsymbol{x}}(t)) \leq f^* + \frac{\|\boldsymbol{x}(0) - \boldsymbol{x}^*\|^2}{2\alpha t} + \frac{\|\boldsymbol{\zeta}(0)\|^2}{2\alpha t} + B_d r + B_{\boldsymbol{\zeta}}\epsilon + \frac{\alpha(B_L+\epsilon)^2}{2} + \frac{\alpha(B_L+r)^2}{2} \tag{25}$$

$$\text{(iii)}\ f(\bar{\boldsymbol{x}}(t)) \geq f^* - B_{\boldsymbol{\zeta}}^*\|[\boldsymbol{g}(\bar{\boldsymbol{x}}(t))]^+\|. \tag{26}$$

In part (i), there are terms which decrease with $t$, and terms which are constant. The constant terms capture the final error level in the constraint violation, while the terms decreasing with $t$ quantify the rate of decrease. Part (ii) provides an upper bound on the difference $f(\bar{\boldsymbol{x}}(t)) - f^*$, and similar comments with part (i) hold regarding the rate of decrease of this bound. In part (iii), the quantity $\|[\boldsymbol{g}(\bar{\boldsymbol{x}}(t))]^+\|$ can be substituted from part (i) to obtain a lower bound for $f(\bar{\boldsymbol{x}}(t)) - f^*$, which also decreases with $t$ in the previously described fashion. It is also worth stressing that apart from the dependence on the iteration index, parts (i)–(iii) provide bounds where the impact of the errors ($\epsilon$, $r$) is explicitly accounted for. It follows further from part (iii) that $f(\bar{\boldsymbol{x}}(t))$ may be smaller than $f^*$, because $\|[\boldsymbol{g}(\bar{\boldsymbol{x}}(t))]^+\|$ might not go to zero. This improvement in the objective value is explained by the infeasibility that is allowed by part (i). Note also that Proposition 1 reduces to [8, Prop. 5.1] with perfect exchanges; i.e., when $r = 0$ and $\epsilon = 0$.

Now define

$$S_t := \sum_{i=0}^{t-1} \beta^{-i}. \tag{27}$$

The result for sequence $\bar{\boldsymbol{x}}_\beta(t)$ corresponding to Proposition 1 is as follows.

**Proposition 2.** *Under Assumptions 1–3, the sequence $\{\bar{\boldsymbol{x}}_\beta(t)\}$ satisfies for $t \geq 2$*

$$\text{(i) } \|[\boldsymbol{g}(\bar{\boldsymbol{x}}_\beta(t))]^+\| \leq \frac{\|\boldsymbol{\zeta}(0)\|^2 + 2\|\boldsymbol{\zeta}(0)\|B_{\boldsymbol{\zeta}} + B_{\boldsymbol{\zeta}}^2}{2S_t\alpha\varrho} + \frac{\|\boldsymbol{x}(0) - \boldsymbol{x}^*\|}{2S_t\alpha\varrho} + \frac{2B_{\boldsymbol{\zeta}}\epsilon + B_d r}{\varrho} + \frac{\alpha(B_L+\epsilon)^2}{2\varrho} + \frac{\alpha(B_L+r)^2}{2\varrho} + \frac{2B_{\boldsymbol{\zeta}}^2 + B_d^2}{2\alpha\varrho}\frac{1-\beta^{t-1}}{1-\beta^t}(1-\beta) \tag{28}$$

$$\text{(ii) } f(\bar{\boldsymbol{x}}_\beta(t)) \leq f^* + \frac{\|\boldsymbol{x}(0) - \boldsymbol{x}^*\|^2}{2\alpha S_t} + \frac{\|\boldsymbol{\zeta}(0)\|^2}{2\alpha S_t} + B_d r + B_{\boldsymbol{\zeta}}\epsilon + \frac{\alpha(B_L+\epsilon)^2}{2} + \frac{\alpha(B_L+r)^2}{2} + \frac{B_{\boldsymbol{\zeta}}^2 + B_d^2}{2\alpha}\frac{1-\beta^{t-1}}{1-\beta^t}(1-\beta) \tag{29}$$

$$\text{(iii) } f(\bar{\boldsymbol{x}}_\beta(t)) \geq f^* - B_{\boldsymbol{\zeta}}^* \|[\boldsymbol{g}(\bar{\boldsymbol{x}}_\beta(t))]^+\|. \tag{30}$$

Similar comments as in Proposition 1 hold for the terms appearing in the right-hand sides of (28)–(30). The dependence on the iteration index, the maximum errors in the gradients, and now, the forgetting factor, is explicit. In these bounds, $S_t$ tends to infinity as $t \to \infty$, so the terms divided by $S_t$ tend to zero. Note that the quantity $(1-\beta^{t-1})/(1-\beta^t)$ is smaller than 1 for all $t \geq 2$, and tends to 1 as $t \to \infty$. Hence, this quantity may be substituted by 1 to obtain simpler bounds in (28)–(30).

Useful insights can be gained by comparing the bounds on the constraint violation in parts (i) of Propositions 1 and 2. Observe that the running averages $\bar{\boldsymbol{x}}_\beta(t)$ achieve an error level exceeding that of $\bar{\boldsymbol{x}}(t)$ by a quantity proportional to $1-\beta$. This difference in the error levels is therefore controlled by the choice of $\beta$, and can be made small because in practice $\beta$ is chosen close to 1. It is also interesting to see that the constraint violation caused by $\bar{\boldsymbol{x}}_\beta(t)$ decreases faster than that of $\bar{\boldsymbol{x}}(t)$. The reason is that the denominator $S_t$ in (28) is exponential in $t$ [cf. (27)] as compared to the denominator $t$ in (24), which is merely linear. Similar comments can be made for the suboptimality in the corresponding parts (ii) and (iii) of the propositions.

Recall that the results so far have been asserted either by assuming that the iterates $\boldsymbol{\zeta}(t)$ are projected on the nonnegative orthant but remain bounded, or, that they are forced to be bounded by projection on a set $\mathcal{D} = \mathcal{D}_{\text{ball}}$ of the form (21) for some fixed $\varrho > 0$. In the latter case, where a projection on a bounded set is chosen, it may be desirable to have a projection on box constraint sets. This is certainly preferable in distributed implementations of the power control algorithms. Note that $\mathcal{D}_{\text{ball}}$ in (21) is a ball, rather than a box. It is possible to consider the following set instead of $\mathcal{D}$, which is a box:

$$\mathcal{D}_{\text{box}} := \{\boldsymbol{\zeta} \geq \mathbf{0} | \ \|\boldsymbol{\zeta}\|_\infty \leq B^*_{\boldsymbol{\zeta}} + \varrho\}. \tag{31}$$

The following corollary asserts that using $\mathcal{D}_{\text{box}}$ for projection, it is possible to obtain performance identical to that claimed by Propositions 1 and 2.

**Corollary 1.** *If $\mathcal{D}_{\text{box}}$ is used instead of $\mathcal{D}_{\text{ball}}$ as the projection set $\mathcal{D}$ in* (15b)*, the results of Propositions 1 and 2 still hold.*

Propositions 1 and 2 along with Corollary 1 suggest that guaranteed performance can be obtained in the power control algorithms impaired by errors in the gradient vectors, if the running averages of the sequences $\{\boldsymbol{y}(t)\}$ and $\{\boldsymbol{z}(t)\}$ are formed.

Note that Propositions 1 and 2 characterize the performance of running averages (16) and (17) when there are errors in the gradient vectors. It is useful to know that if there are no errors, the performance achieved by the running averages of the iterates is the same as the one achieved by the iterates themselves. In the context of power control algorithms, this performance when there are no errors in the gradients, is given by Lemma 3. This lemma ensures that the iterates will remain arbitrarily close to the set of optimal primal solutions for appropriate choice of the stepsize. In particular, consider choosing an $\varepsilon$ and the appropriate stepsize $\alpha$ so that the iterates $\{\boldsymbol{x}(t)\}$ satisfy $\text{dist}(\boldsymbol{x}(t), \mathcal{X}^*) \leq \varepsilon$ for all $t \geq t'$, as Lemma 3 asserts. The following lemma states that the running averages $\{\bar{\boldsymbol{x}}(t)\}$ and $\{\bar{\boldsymbol{x}}_\beta(t)\}$ will asymptotically have distance at most $\varepsilon$ from the optimal $\mathcal{X}^*$, and is reminiscent of Toeplitz's lemma [25, p. 341].

**Lemma 4.** *Suppose the sequence $\{\boldsymbol{x}(t)\}$ in $\mathbb{R}^N$ satisfies $\text{dist}(\boldsymbol{x}(t), \mathcal{X}^*) \leq \varepsilon$ for all $t \geq t'$, where $t'$ is a given integer, and $\mathcal{X}^*$ is convex, closed, and bounded. Then, any limit point (for instance, call one $\mathring{\boldsymbol{x}}$) of* *sequence* $\{\bar{\boldsymbol{x}}(t)\}$ *or* $\{\bar{\boldsymbol{x}}_\beta(t)\}$ *satisfies* $\operatorname{dist}(\mathring{\boldsymbol{x}}, \mathcal{X}^*) \leq \varepsilon$.

Lemma 4 suggests that one can use $\bar{\boldsymbol{x}}(t)$ or $\bar{\boldsymbol{x}}_\beta(t)$ instead of $\boldsymbol{x}(t)$ and still enjoy $\varepsilon$-optimal convergence of average power control iterates.

### *C. Selecting Convergence Parameters*

Several quantities are introduced in the previous subsections under Assumptions 1–3. Their knowledge is useful for running (15)—for instance, compact sets $\mathcal{X}$ and $\mathcal{D}$—as well as for the performance analysis. In what follows, certain guidelines for choosing those quantities are provided in the power control context of Section II. The focus here is on the single-channel formulations under C1–C3.

- In order to have a compact set $\mathcal{X}$, it is necessary to know bounds on the optimal powers $p_i^*$ and the optimal local IpN variables $q_i^*$ (when applicable). First, it should be stressed that the optimal powers are non-zero, due to C1. Hence, constants $p_i^{\min} > 0$ are *guaranteed* to exist so that

$$p_i^{\min} \leq p_i^* \leq p_i^{\max}, \quad \forall i \in \mathcal{M}. \tag{32}$$

  In practice, $p_i^{\min}$ can be selected to be a very small positive number.
- If user $i$ has a minimum SINR constraint $\gamma_i^{\min}$, then any $p_i^{\min}$ satisfying

$$\frac{h_{ii} p_i^{\min}}{n_i} \leq \gamma_i^{\min} \tag{33}$$

  can be used as a bound in (32). Indeed, if $p_i^* < p_i^{\min}$ for some $p_i^{\min}$ satisfying (33), it holds that

$$\frac{h_{ii} p_i^*}{n_i + \sum_{k \neq i} h_{ki} p_k^*} < \frac{h_{ii} p_i^{\min}}{n_i} < \gamma_i^{\min} \tag{34}$$

  which contradicts the optimality of $p_i^*$.
- Regarding $q_i^*$ [cf. (4)], the optimality of the relaxation implies that

$$n_i + \sum_{k \neq i} h_{ki} p_k^{\min} \leq q_i^* \leq n_i + \sum_{k \neq i} h_{ki} p_k^{\max} \quad \forall i \in \mathcal{M}. \tag{35}$$

- An explicit bound on the optimal Lagrange multipliers calls for a lower bound $\check{d}$ on the optimal dual value of problem (9) [cf. (22)]. To this end, define

$$\breve{\gamma}_i := \frac{h_{ii} p_i^{\max}}{n_i}. \tag{36}$$

  It is clear that for problem (2) and any feasible $\boldsymbol{p}$, it holds that $\breve{\gamma}_i > h_{ii} p_i / (n_i + \sum_{k \neq i} h_{ki} p_k)$ for all $i \in \mathcal{M}$. Similarly, for (4) and any feasible $\boldsymbol{p}$ and $\boldsymbol{q}$, it holds that $\breve{\gamma}_i > h_{ii} p_i / q_i$ for all $i \in \mathcal{M}$.

Therefore, since the utilities $u_i(.)$ are selected strictly increasing and the duality gap is zero, it suffices to pick any $\check{d}$ satisfying

$$\check{d} < -\sum_{i=1}^{M} u_i(\check{\gamma}_i) < f(\boldsymbol{x}^*) = f^*. \tag{37}$$

- A vector $\check{\boldsymbol{x}}$ satisfying Slater's constraint qualification is also needed.
  1) For problem (2) with only the constraints (2c) or (2d) imposed, sufficiently small powers can be selected so that these constraints hold as strict inequalities.
  2) For problems including SINR constraints, the power control algorithm of [23] can be applied to return powers achieving given SINR constraints if the latter are feasible. Hence, the algorithm can be applied for (2a)–(2b) to any vector of SINRs $\check{\boldsymbol{\gamma}}$ with $\gamma_i^{\min} < \check{\gamma}_i$ for all $i \in \mathcal{M}$. If those SINRs (or any other strictly greater than $\gamma_i^{\min}$ for all $i$) are achievable with powers $\check{\boldsymbol{p}}$, then $\check{\boldsymbol{p}}$ is the desired Slater vector. For Applications 1 and 3, the algorithm can be applied to any vector of SINRs $\check{\boldsymbol{\gamma}}$ with $\gamma_i^{\min} < \check{\gamma}_i < \gamma_i^{\max}$ for all $i \in \mathcal{M}$. If these are achievable with powers $\check{\boldsymbol{p}}$, variables $\check{q}_i$ can be chosen with this $\check{\boldsymbol{p}}$ so that $\check{\gamma}_i < h_{ii}\check{p}_i/\check{q}_i < \gamma_i^{\max}$ for all $i \in \mathcal{M}$.

## IV. Numerical Tests

Distributed power control for the flexible primary model with interference protection (cf. Application 2) and imperfect exchanges is tested numerically here. There are $M = 10$ users; the set of PUs is $\mathcal{M}_p = \{1, 2, 5, 6, 8\}$, and the set of SUs is $\mathcal{M}_s = \{3, 4, 7, 9, 10\}$. The locations of Tx-Rx pairs are given in Table I. With $d_{ij}$ denoting the distance between $\text{Tx}_i$ and $\text{Rx}_j$, the channels follow the models $h_{ii} = 0.5d_{ij}^{-2.5}$ and $h_{ij} = 0.05d_{ij}^{-2.5}$ for $i \neq j$. The rest of the parameters are listed in Table II. Errors are considered only in the updates of $y_i$. The errors are random i.i.d. across iterations and across users, uniformly distributed with mean -0.005 and support 0.08. This model gives relatively significant errors, with value of $r$ (cf. Assumption 3) three orders of magnitude larger than the max-norm of the gradient vector at the optimal point. (This vector is easily obtained by running algorithm (11).)

It was observed that the primal iterates as well as the Lagrange multiplier iterates were indeed bounded, so Assumption 2 was satisfied. The interference and SINR constraints as well as the results are presented in Table III. For the algorithm with errors in the gradients, the running averages $\bar{\boldsymbol{y}}_\beta(t)$ are used for the evaluation of the achieved SINRs and interferences in Table III. It is seen that the constraints are accurately met.

Fig. 1 shows the constraint violation for the two types of averaging (16) and (17). Recall that the function $\boldsymbol{g}(\boldsymbol{x})$ is the vector formed by stacking the constraints (4b) and (4c) for all $i \in \mathcal{M}$. It is observed

TABLE I

COORDINATES OF 10 TX-RX PAIRS IN METERS (SHOWN IN 2 COLUMNS). TX ARE DEPLOYED OVER A SQUARE AREA OF SIDE 30M. EACH RX IS LOCATED IN A SQUARE AREA OF SIDE 5M CENTERED AT ITS PEER TX, AND AT LEAST 1M AWAY FROM ITS PEER TX. POSITIONS WERE RANDOMLY SELECTED.

| $\text{Tx}_i$; $\text{Rx}_i$ $(i = 1, 2, 3, 4, 5)$ | $\text{Tx}_i$; $\text{Rx}_i$ $(i = 6, 7, 8, 9, 10)$ |
|---|---|
| (21.181, 21.061); (19.755, 20.88) | (0.91656, 3.5594); (2.7782, 2.8003) |
| (20.508, 0.53543); (21.837, -1.2724) | (0.61918, 1.8744); (-1.6297, 1.3382) |
| ( 4.6367, 16.308); (3.6544, 14.561) | (9.152, 0.67406); (11.408, 0.99965) |
| ( 5.3579, 17.061); (5.2071, 18.869) | (1.7077, 24.352); (0.10774, 21.941) |
| (12.942, 10.246); (10.602, 8.0944) | (28.384, 22.046); (28.627, 23.873) |

TABLE II

SIMULATION PARAMETERS FOR ALGORITHM WITH IMPERFECT EXCHANGES.

| |
|---|
| $M = 10$, $\mathcal{M}_p = \{1, 2, 5, 6, 8\}$, $\mathcal{M}_s = \{3, 4, 7, 9, 10\}$ |
| $u_i(\gamma_i) = w_i \ln \gamma_i$; $w_i = 1$, $i \in \mathcal{M}_p$; $w_i = 0.5$, $i \in \mathcal{M}_s$ |
| $p_i^{\max} = 6$ dBm, $i \in \mathcal{M}_p$; $p_i^{\max} = 0$ dBm, $i \in \mathcal{M}_s$ |
| $n_i = -41$ dBm, for all $i \in \mathcal{M}$ |
| Initialization: $p_i^{\max} = 1\%$ of $p_i^{\max}$, $z_i = \ln n_i$ for all $i \in \mathcal{M}$<br>$\mu_i = 1$, for all $i \in \mathcal{M}$; $\lambda_i = 0$, $\nu_i = 0$ for all $i \in \mathcal{M}_s$ |
| $\alpha = 0.06$, $\beta = 0.9$, $\mathcal{D} =$ nonnegative orthant |

that the constraint violation for the averaging (17) decreases faster than that for (16).

## V. CONCLUSIONS

This paper presented in a unified way several convex power control formulations, ranging from typical to contemporary ones pertaining to spectrum sharing in unlicensed bands and bands with primary and secondary users. The common feature is that the power control must account for various constraints which couple the power variables, such as maximum interference constraints, maximum receive-power constraints, or minimum and maximum SINR constraints. The gradient method for finding saddle points of the associated Lagrangian function was employed. Distributed implementations of the method were also presented in a unified way, by identifying the non-local parts of the gradient vectors for each link. In order to acquire these, exchange of information among nodes takes place, which may entail errors in the gradient vectors.

The impact of errors in the gradient method for saddle points was studied in a general setting. To

TABLE III

POWER CONTROL WITH IMPERFECT EXCHANGES: SUM-UTILITY (TOP); CONSTRAINT VALUES AND ACHIEVED SINR AND INTERFERENCE PER USER WITH AND WITHOUT ERRORS (BOTTOM). THE FIRST GROUP OF USERS CONSISTS OF PUS, WITH MAXIMUM INTERFERENCE CONSTRAINTS, AND THE SECOND GROUP OF SUS, WITH MINIMUM (LEFT) AND MAXIMUM (RIGHT) SINR CONSTRAINTS.

| | | Constraints | | Without errors | With errors |
|---|---|---|---|---|---|
| | Sum-utility | | | 26.635 | 26.627 |
| 1 | Interference (dBm) | -30 | | -38.402 | -38.392 |
| 2 | | -30 | | -39.684 | -39.679 |
| 5 | | -38 | | -38.132 | -38.129 |
| 6 | | -38 | | -37.997 | -37.995 |
| 8 | | -38 | | -38.002 | -37.995 |
| 3 | SINR (dB) | 0 | 10 | 10.000 | 10.040 |
| 4 | | 0 | 10 | 10.000 | 9.958 |
| 7 | | 0 | 20 | -0.001 | 0.012 |
| 9 | | 0 | 10 | 10.00 | 9.975 |
| 10 | | 0 | 20 | 20.00 | 19.989 |

this end, two running averages (ergodic sequences) were formed from the iterates generated by the method. One weighs all iterates the same, and the other uses exponentially decreasing weights for past iterates. Under the assumption of bounded—but otherwise arbitrary—errors, the two averaging schemes were shown to have complementary strengths in terms of constraint violation and suboptimality reduction. Future research could focus on characterizing the impact of random errors arising from probability density functions with infinite support, using stochastic approximation techniques.

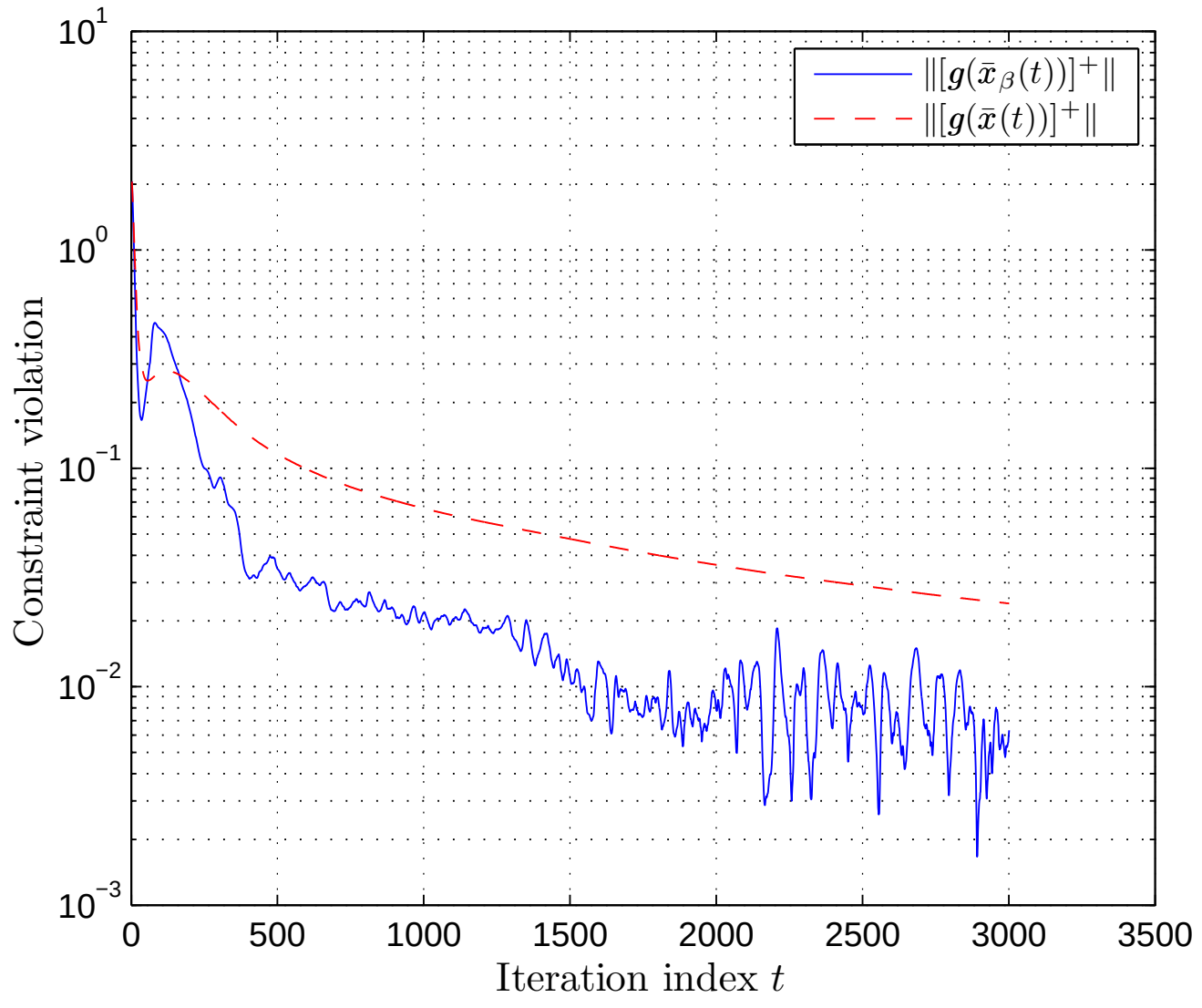

Fig. 1. Constraint violation for two types of averaging [cf. (16) and (17)].

## References

[1] S. Stańczak, A. Feistel, H. Boche, and M. Wiczanowski, "Towards efficient and fair resource allocation in wireless networks," in *Proc. 6th Int. Symp. Modeling and Optimization in Mobile, Ad Hoc, and Wireless Networks*, Berlin, Germany, Mar.–Apr. 2008.

[2] D. Julian, M. Chiang, D. O'Neill, and S. Boyd, "QoS and fairness constrained convex optimization of resource allocation for wireless cellular and ad hoc networks," in *Proc. 21st IEEE INFOCOM Conf.*, New York, NY, June 2002, pp. 477–486.

[3] M. Wiczanowski, S. Stańczak, and H. Boche, "Performance and interference control in wireless ad hoc and mesh networks—a generalized Lagrangian approach," *IEEE Trans. Signal Process.*, vol. 56, no. 8, pp. 4039–4052, Aug. 2008.

[4] N. Gatsis, A. G. Marques, and G. B. Giannakis, "Power control for cooperative dynamic spectrum access networks with diverse QoS constraints," *IEEE Trans. Commun.*, vol. 58, no. 3, pp. 933–944, Mar. 2010.

[5] C. Zou, T. Jin, C. Chigan, and Z. Tian, "QoS-aware distributed spectrum sharing for heterogeneous wireless cognitive networks," *Comput. Netw.*, vol. 52, no. 4, pp. 864–878, Mar. 2008.

[6] D. P. Bertsekas and J. N. Tsitsiklis, *Parallel and Distributed Computation: Numerical Methods*. Belmont, MA: Athena Scientific, 1997.

[7] D. Maistrovskii, "Gradient methods for finding saddle points," *Matekon*, vol. 14, no. 1, pp. 3–22, 1977.

[8] A. Nedić and A. Ozdaglar, "Subgradient methods for saddle point problems," *J. Optim. Theory Appl.*, vol. 142, no. 1, pp. 205–228, July 2009.

[9] S. Stańczak, M. Wiczanowski, and H. Boche, *Fundamentals of Resource Allocation in Wireless Networks: Theory and Algorithms*, ser. Foundations in Signal Processing, Communications and Networking. Berlin, Germany: Springer-Verlag, 2009, vol. 3.

[10] M. Chiang, “Balancing transport and physical layers in wireless multihop networks: Jointly optimal congestion control and power control,” *IEEE J. Select. Areas Commun.*, vol. 23, no. 1, pp. 104–116, Jan. 2005.

[11] H. J. Kushner and D. S. Clark, *Stochastic Approximation Methods for Constrained and Unconstrained Systems*. New York, NY: Springer-Verlag, 1978.

[12] R. Buche and H. J. Kushner, “Rate of convergence for constrained stochastic approximation algorithms,” *SIAM J. Control Optim.*, vol. 40, no. 4, pp. 1011–1041, 2001.

[13] M. V. Solodov, “Convergence analysis of perturbed feasible descent methods,” *J. Optim. Theory Appl.*, vol. 93, no. 2, pp. 337–353, May 1997.

[14] A. Nedić and D. P. Bertsekas, “The effect of determinist noise in subgradient methods,” *Math. Progr.*, 2009. [Online]. Available: http://dx.doi.org/10.1007/s10107-008-0262-5

[15] S. S. Ram, A. Nedić, and V. V. Veeravalli, “Distributed stochastic subgradient projection algorithms for convex optimization,” 2008, submitted for publication. [Online]. Available: http://arxiv.org/abs/0811.2595

[16] B. T. Polyak, “New method of stochastic approximation type,” *Autom. Remote Control*, vol. 51, no. 7, pp. 937–946, July 1990.

[17] ——, *Introduction to Optimization*. New York, NY: Optimization Software, 1987.

[18] S. Stańczak, M. Wiczanowski, and H. Boche, “Distributed utility-based power control: Objectives and algorithms,” *IEEE Trans. Signal Process.*, vol. 55, no. 10, pp. 5058–5068, Oct. 2007.

[19] H. Iiduka and I. Yamada, “An ergodic algorithm for the power-control games for CDMA data networks,” *J. Math. Model. Alg.*, vol. 8, no. 1, pp. 1–18, Mar. 2009.

[20] J. M. Peha, “Approaches to spectrum sharing,” *IEEE Commun. Mag.*, vol. 43, no. 2, pp. 10–12, Feb. 2005.

[21] P. Hande, S. Rangan, M. Chiang, and X. Wu, “Distributed uplink power control for optimal SIR assignment in cellular data networks,” *IEEE/ACM Trans. Netw.*, vol. 16, no. 6, pp. 1420–1433, Dec. 2008.

[22] S. Boyd and L. Vandenberghe, *Convex Optimization*. Cambridge, UK: Cambridge University Press, 2004.

[23] R. D. Yates, “A framework for uplink power control in cellular radio systems,” *IEEE J. Select. Areas Commun.*, vol. 13, no. 7, pp. 1341–1347, Sep. 1995.

[24] D. P. Bertsekas, A. Nedić, and A. E. Ozdaglar, *Convex Analysis and Optimization*. Belmont, MA: Athena Scientific, 2003.

[25] V. Solo and X. Kong, *Adaptive Signal Processing Algorithms: Stability and Performance*. Upper Saddle River, NJ: Prentice Hall, 1995.

## APPENDIX A

## PROOFS OF LEMMAS 1 AND 2

*Proof of Lemma 1:* First, Lemma 1 is proved in two steps. Define $\boldsymbol{\gamma}^{\max} := [\gamma_1^{\max}, \dots, \gamma_M^{\max}]^T$.

*Step 1:* The following claim is proved, which applies to the case where all users have maximum SINR constraints in (3). It relates the (non-)achievability of $\boldsymbol{\gamma}^{\max}$ in problem (3) (cf. C3) and the relaxed (4).

*Claim:* If there is no $\boldsymbol{p}$ in the feasible set of (3) such that $\gamma_i = \gamma_i^{\max}$ for all $i \in \mathcal{M}$, then there are no $\boldsymbol{p}$, $\boldsymbol{q}$ in the feasible set of (4) such that $h_{ii}p_i/q_i = \gamma_i^{\max}$ for all $i \in \mathcal{M}$.

With $\boldsymbol{\iota}^{\max} := [q_1^{\max}/h_{11}, \ldots, q_M^{\max}/h_{MM}]^T$, feasibility of the SINRs $\gamma_i^{\max}$ in (3) is written as

$$\boldsymbol{p} = \mathbf{D}(\boldsymbol{\gamma}^{\max})\mathbf{A}\boldsymbol{p} + \mathbf{D}(\boldsymbol{\gamma}^{\max})\boldsymbol{\eta}, \quad \mathbf{0} < \boldsymbol{p} \leq \boldsymbol{p}^{\max}, \quad \mathbf{A}\boldsymbol{p} + \boldsymbol{\eta} \leq \boldsymbol{\iota}^{\max}. \tag{38}$$

If the spectral radius of $\mathbf{D}(\boldsymbol{\gamma}^{\max})\mathbf{A}$ satisfies $\rho(\mathbf{D}(\boldsymbol{\gamma}^{\max})\mathbf{A}) < 1$, then the first linear system in (38) has a unique positive solution $\boldsymbol{p}(\boldsymbol{\gamma}^{\max}) := (\mathbf{I} - \mathbf{D}(\boldsymbol{\gamma}^{\max})\mathbf{A})^{-1}\mathbf{D}(\boldsymbol{\gamma}^{\max})\boldsymbol{\eta}$ [9, Th. A.51]. Since it is assumed that (38) does not have a solution, one of the following two mutually exclusive cases can happen: (i) $\rho(\mathbf{D}(\boldsymbol{\gamma}^{\max})\mathbf{A}) \geq 1$; or (ii) $\rho(\mathbf{D}(\boldsymbol{\gamma}^{\max})\mathbf{A}) < 1$ but with $\boldsymbol{p}(\boldsymbol{\gamma}^{\max}) \nleq \boldsymbol{p}^{\max}$ or $\mathbf{A}\boldsymbol{p}(\boldsymbol{\gamma}^{\max}) + \boldsymbol{\eta} \nleq \boldsymbol{\iota}^{\max}$. (The notation $\nleq$ means that at least one entry of the vectors does not satisfy the inequality $\leq$.) The cases of $\rho(\mathbf{D}(\boldsymbol{\gamma}^{\max})\mathbf{A}) \geq 1$ or of $\rho(\mathbf{D}(\boldsymbol{\gamma}^{\max})\mathbf{A}) < 1$ with $\boldsymbol{p}(\boldsymbol{\gamma}^{\max}) \nleq \boldsymbol{p}^{\max}$ are dealt with in [4, Lemma 3].

To address the remaining case, note that achievability of $\boldsymbol{\gamma}^{\max}$ in (4) implies that $\mathbf{D}^{-1}(\boldsymbol{\gamma}^{\max})\boldsymbol{p} \leq \boldsymbol{\iota}^{\max}$. In order for such $\boldsymbol{p}$ to exist, it is necessary that $\mathbf{D}^{-1}(\boldsymbol{\gamma}^{\max})\boldsymbol{p}(\boldsymbol{\gamma}^{\max}) \leq \boldsymbol{\iota}^{\max}$. But using $\rho(\mathbf{D}(\boldsymbol{\gamma}^{\max})\mathbf{A}) < 1$ and $\mathbf{A}\boldsymbol{p}(\boldsymbol{\gamma}^{\max}) + \boldsymbol{\eta} \nleq \boldsymbol{\iota}^{\max}$, it follows from the first linear system in (38) that $\mathbf{D}^{-1}(\boldsymbol{\gamma}^{\max})\boldsymbol{p}(\boldsymbol{\gamma}^{\max}) = \mathbf{A}\boldsymbol{p}(\boldsymbol{\gamma}^{\max}) + \boldsymbol{\eta} \nleq \boldsymbol{\iota}^{\max}$ (contradiction).

*Step 2:* Now, (5) is proved by contradiction. The previous claim is used, and the proof follows the proof of [4, Prop. 1], noting that the additional constraint $q_i \leq q_i^{\max}$ can be easily taken into account.

Finally, the multi-channel case does not need a claim similar to the one in Step 1, but follows the proof of [4, Prop. 3], where again the additional constraint $\boldsymbol{q}_i \in \mathcal{Q}_i$ can be easily taken into account. □

*Proof of Lemma 2:* The proof is for the single-channel case first, and proceeds in two steps.

*Step 1:* The aim is to show that $\mu_i^* > 0$, $i = 1, \ldots, M$, where $\mu_i^*$ are the optimal Lagrange multipliers corresponding to constraint (4c). To this end, note that the additional convex constraint set $\mathcal{Y} \times \mathcal{Z}$ has special structure, described by the following property: If $(\boldsymbol{y}, \boldsymbol{z}) \in \mathcal{Y} \times \mathcal{Z}$, then any $(\breve{\boldsymbol{y}}, \breve{\boldsymbol{z}})$ with $\breve{\boldsymbol{y}} \leq \boldsymbol{y}$ and $\breve{\boldsymbol{z}} \leq \boldsymbol{z}$ also satisfies $(\breve{\boldsymbol{y}}, \breve{\boldsymbol{z}}) \in \mathcal{Y} \times \mathcal{Z}$. This property, together with Slater's constraint qualification, can be used to write the necessary optimality conditions [24, Sec. 5.2 and 5.4]

$$\left.\frac{\partial L(\boldsymbol{\omega})}{\partial y_i}\right|_{(\boldsymbol{y}^*, \boldsymbol{z}^*, \boldsymbol{\nu}^*, \boldsymbol{\lambda}^*, \boldsymbol{\mu}^*)} \leq 0, \quad \left.\frac{\partial L(\boldsymbol{\omega})}{\partial z_i}\right|_{(\boldsymbol{y}^*, \boldsymbol{z}^*, \boldsymbol{\nu}^*, \boldsymbol{\lambda}^*, \boldsymbol{\mu}^*)} \leq 0 \quad \forall i \in \mathcal{M}. \tag{39}$$

Writing out the partial derivatives explicitly and summing them up, we arrive at a linear system of equations that the vector $\boldsymbol{\mu}^* := [\mu_1^*, \ldots, \mu_M^*]$ satisfies which can be shown to have a positive solution, in the spirit of the proof of [4, Lemma 1(a)].

*Step 2:* The positivity of the optimal Lagrange multipliers for the local IpN constraints can be leveraged to show that the Hessian of the Lagrangian function with respect to $(\boldsymbol{y}, \boldsymbol{z})$ is stritly convex [4, Lemma 1(b)]. This implies stability of the saddle points with respect to $(\boldsymbol{y}, \boldsymbol{z})$.

Finally, the multi-channel case uses analogous arguments, noting that the additional convex constraint set under both types of interference requirements has the property previously described. □

## APPENDIX B
## PROOFS OF PROPOSITIONS 1 AND 2

Assumptions 1–3—referred to as "boundedness assumptions" hereafter—are supposed to hold throughout. Let $\boldsymbol{x}^*$ and $\boldsymbol{\zeta}^*$ denote an optimal primal solution and an optimal Lagrange multiplier vector for (9), respectively. For the analysis, the stability of the saddle points [cf. (12)] will not be used, but instead methods from [8] will be adapted in order to (a) include errors in the updates, and (b) accomodate the time-varying weights $\beta^{-i}$. A lemma about the successive iterates $\boldsymbol{x}(t)$, $\boldsymbol{\zeta}(t)$ and $\boldsymbol{x}(t+1)$, $\boldsymbol{\zeta}(t+1)$ generated by (15) follows, which is useful in proving results for both types of averaging.

**Lemma 5.** *Under Assumptions 1–3, the sequences* $\{\boldsymbol{x}(t)\}$, $\{\boldsymbol{\zeta}(t)\}$ *satisfy for all* $\boldsymbol{x} \in \mathcal{X}$, $\boldsymbol{\zeta} \in \mathcal{D}$, *and* $t \geq 0$

$$\text{(i)}\ \|\boldsymbol{\zeta}(t+1)-\boldsymbol{\zeta}\|^2 \leq \|\boldsymbol{\zeta}(t)-\boldsymbol{\zeta}\|^2 + 2\alpha\nabla_{\boldsymbol{\zeta}}^T L(\boldsymbol{x}(t),\boldsymbol{\zeta}(t))(\boldsymbol{\zeta}(t)-\boldsymbol{\zeta}) + 2\alpha\epsilon\|\boldsymbol{\zeta}(t)-\boldsymbol{\zeta}\| + \alpha^2(B_L+\epsilon)^2 \quad (40)$$

$$\text{(ii)}\ \|\boldsymbol{x}(t+1)-\boldsymbol{x}\|^2 \leq \|\boldsymbol{x}(t)-\boldsymbol{x}\|^2 - 2\alpha(L(\boldsymbol{x}(t),\boldsymbol{\zeta}(t)) - L(\boldsymbol{x},\boldsymbol{\zeta}(t))) + 2\alpha r\|\boldsymbol{x}(t)-\boldsymbol{x}\| + \alpha^2(B_L+r)^2. \quad (41)$$

*Proof of Lemma 5:* (i) Using the nonexpansive property of the projection [24, Prop. 2.2.1], it follows from (15b) that for all $\boldsymbol{x} \in \mathcal{X}$,

$$\|\boldsymbol{\zeta}(t+1)-\boldsymbol{\zeta}\|^2 \leq \|\boldsymbol{\zeta}(t) + \alpha(\nabla_{\boldsymbol{\zeta}} L(\boldsymbol{x}(t),\boldsymbol{\zeta}(t)) + \boldsymbol{\epsilon}(t)) - \boldsymbol{\zeta}\|^2$$
$$= \|\boldsymbol{\zeta}(t)-\boldsymbol{\zeta}\|^2 + 2\alpha\nabla_{\boldsymbol{\zeta}}^T L(\boldsymbol{x}(t),\boldsymbol{\zeta}(t))(\boldsymbol{\zeta}(t)-\boldsymbol{\zeta}) + 2\alpha\boldsymbol{\epsilon}^T(t)(\boldsymbol{\zeta}(t)-\boldsymbol{\zeta}) + \alpha^2\|\nabla_{\boldsymbol{\zeta}} L(\boldsymbol{x}(t),\boldsymbol{\zeta}(t)) + \boldsymbol{\epsilon}(t)\|^2.$$

Invoking Cauchy-Schwartz inequality and the boundedness assumptions, (40) follows readily.

(ii) Again, using the nonexpansive property of the projection, it follows from (15a) that for all $\boldsymbol{\zeta} \in \mathcal{D}$,

$$\|\boldsymbol{x}(t+1)-\boldsymbol{x}\|^2 \leq \|\boldsymbol{x}(t) + \alpha(\nabla_{\boldsymbol{x}} L(\boldsymbol{x}(t),\boldsymbol{\zeta}(t)) + \boldsymbol{r}(t)) - \boldsymbol{x}\|^2$$
$$= \|\boldsymbol{x}(t)-\boldsymbol{x}\|^2 + 2\alpha\nabla_{\boldsymbol{\zeta}}^T L(\boldsymbol{x}(t),\boldsymbol{\zeta}(t))(\boldsymbol{x}(t)-\boldsymbol{x}) + 2\alpha\boldsymbol{r}^T(t)(\boldsymbol{x}(t)-\boldsymbol{x}) + \alpha^2\|\nabla_{\boldsymbol{x}} L(\boldsymbol{x}(t),\boldsymbol{\zeta}(t)) + \boldsymbol{r}(t)\|^2.$$

Invoking Cauchy-Schwartz inequality and the boundedness assumptions, it is deduced that

$$\|\boldsymbol{x}(t+1)-\boldsymbol{x}\|^2 \leq \|\boldsymbol{x}(t)-\boldsymbol{x}\|^2 + 2\alpha\nabla_{\boldsymbol{\zeta}}^T L(\boldsymbol{x}(t),\boldsymbol{\zeta}(t))(\boldsymbol{x}(t)-\boldsymbol{x}) + 2\alpha r\|\boldsymbol{x}(t)-\boldsymbol{x}\| + \alpha^2\|B_L+r\|^2.$$

By convexity of $L(\boldsymbol{x},\boldsymbol{\zeta})$ with respect to $\boldsymbol{x}$, it follows that

$$\nabla_{\boldsymbol{\zeta}}^T L(\boldsymbol{x}(t),\boldsymbol{\zeta}(t))(\boldsymbol{x}(t)-\boldsymbol{x}) + L(\boldsymbol{x}(t),\boldsymbol{\zeta}(t)) \leq L(\boldsymbol{x},\boldsymbol{\zeta}(t)).$$

Combining the last two inequalities yields (41). □

Now, the focus turns to averaging (16). To facilitate proving the results, we will also consider the running average $\bar{\boldsymbol{\zeta}}(t)$ of Lagrange multipliers, together with $\bar{\boldsymbol{x}}(t)$:

$$\bar{\boldsymbol{\zeta}}(t) := \frac{1}{t}\sum_{i=0}^{t-1}\boldsymbol{\zeta}(i). \tag{42}$$

The following lemma characterizes the running average of the Lagrangian function values at $\boldsymbol{x}(t)$ and $\boldsymbol{\zeta}(t)$, and will be used in the proof of Proposition 1.

**Lemma 6.** *Under Assumptions 1–3, it holds for all $t \geq 1$ that*

$$\frac{1}{t}\sum_{i=0}^{t-1} L(\boldsymbol{x}(i),\boldsymbol{\zeta}(i)) - f^* \leq \frac{\|\boldsymbol{x}(0)-\boldsymbol{x}^*\|^2}{2\alpha t} + rB_d + \frac{\alpha(B_L+r)^2}{2}. \tag{43}$$

*Proof of Lemma 6:* It follows from (41) that for all $\boldsymbol{x} \in \mathcal{X}$, $i \geq 0$

$$L(\boldsymbol{x}(i),\boldsymbol{\zeta}(i)) - L(\boldsymbol{x},\boldsymbol{\zeta}(i)) \leq \frac{\|\boldsymbol{x}(i)-\boldsymbol{x}\|^2 - \|\boldsymbol{x}(i+1)-\boldsymbol{x}\|^2}{2\alpha} + r\|\boldsymbol{x}(i)-\boldsymbol{x}\| + \frac{\alpha(B_L+r)^2}{2}. \tag{44}$$

Averaging the latter, it follows that

$$\frac{1}{t}\sum_{i=0}^{t-1} L(\boldsymbol{x}(i),\boldsymbol{\zeta}(i)) - \frac{1}{t}\sum_{i=0}^{t-1} L(\boldsymbol{x},\boldsymbol{\zeta}(i)) \leq \frac{\|\boldsymbol{x}(i)-\boldsymbol{x}\|^2}{2\alpha t} + \frac{1}{t}\sum_{i=0}^{t-1} r\|\boldsymbol{x}(i)-\boldsymbol{x}\| + \frac{\alpha(B_L+r)^2}{2}. \tag{45}$$

Furthermore, the concavity of $L(\boldsymbol{x},\boldsymbol{\zeta})$ with respect to $\boldsymbol{\zeta}$ implies that

$$\frac{1}{t}\sum_{i=0}^{t-1} L(\boldsymbol{x}^*,\boldsymbol{\zeta}(i)) \geq L(\boldsymbol{x}^*,\bar{\boldsymbol{\zeta}}(t)) \geq L(\boldsymbol{x}^*,\boldsymbol{\zeta}^*) = f^*. \tag{46}$$

Using the latter, $\boldsymbol{x} = \boldsymbol{x}^*$, and the boundedness assumptions into (45), we obtain (43). □

*Proof of Proposition 1:* (i) It follows from (40) for all $\boldsymbol{\zeta} \in \mathcal{D}$ and $i \geq 0$ that

$$\nabla_{\boldsymbol{\zeta}}^T L(\boldsymbol{x}(i),\boldsymbol{\zeta}(i))(\boldsymbol{\zeta}-\boldsymbol{\zeta}(i)) \leq \frac{\|\boldsymbol{\zeta}(i)-\boldsymbol{\zeta}\|^2 - \|\boldsymbol{\zeta}(i+1)-\boldsymbol{\zeta}\|^2}{2\alpha} + \epsilon\|\boldsymbol{\zeta}(i)-\boldsymbol{\zeta}\| + \frac{\alpha(B_L+\epsilon)^2}{2}. \tag{47}$$

Using the convexity of $L(\boldsymbol{x},\boldsymbol{\zeta})$ with respect to $\boldsymbol{\zeta}$, we obtain

$$(\boldsymbol{\zeta}(i)-\boldsymbol{\zeta}^*)^T\nabla_{\boldsymbol{\zeta}} L(\boldsymbol{x}(i),\boldsymbol{\zeta}(i)) \leq L(\boldsymbol{x}(i),\boldsymbol{\zeta}(i)) - L(\boldsymbol{x}(i),\boldsymbol{\zeta}^*) \leq L(\boldsymbol{x}(i),\boldsymbol{\zeta}(i)) - f^*. \tag{48}$$

Upon combining (47) and (48), one arrives at

$$(\boldsymbol{\zeta}-\boldsymbol{\zeta}^*)^T\nabla_{\boldsymbol{\zeta}} L(\boldsymbol{x}(i),\boldsymbol{\zeta}(i)) \leq (\boldsymbol{\zeta}-\boldsymbol{\zeta}(i))^T\nabla_{\boldsymbol{\zeta}} L(\boldsymbol{x}(i),\boldsymbol{\zeta}(i)) + (\boldsymbol{\zeta}(i)-\boldsymbol{\zeta}^*)^T\nabla_{\boldsymbol{\zeta}} L(\boldsymbol{x}(i),\boldsymbol{\zeta}(i)) \tag{49}$$

$$\leq \frac{\|\boldsymbol{\zeta}(i)-\boldsymbol{\zeta}\|^2 - \|\boldsymbol{\zeta}(i+1)-\boldsymbol{\zeta}\|^2}{2\alpha} + \epsilon\|\boldsymbol{\zeta}(i)-\boldsymbol{\zeta}\| + \frac{\alpha(B_L+\epsilon)^2}{2} + L(\boldsymbol{x}(i),\boldsymbol{\zeta}(i)) - f^*. \tag{50}$$

Averaging the latter and invoking (43), we obtain for all $\boldsymbol{\zeta} \in \mathcal{D}$

$$\frac{1}{t}\sum_{i=0}^{t-1}(\boldsymbol{\zeta}-\boldsymbol{\zeta}^*)^T\nabla_{\boldsymbol{\zeta}} L(\boldsymbol{x}(i),\boldsymbol{\zeta}(i)) \leq \frac{\|\boldsymbol{\zeta}(0)-\boldsymbol{\zeta}\|^2}{2\alpha t} + \frac{1}{t}\sum_{i=0}^{t-1}\epsilon\|\boldsymbol{\zeta}(i)-\boldsymbol{\zeta}\| + \frac{\alpha(B_L+\epsilon)^2}{2}$$
$$+ \frac{\|\boldsymbol{x}(0)-\boldsymbol{x}^*\|^2}{2\alpha t} + rB_d + \frac{\alpha(B_L+r)^2}{2}. \tag{51}$$

Now, consider the following vector:

$$\breve{\boldsymbol{\zeta}} := \boldsymbol{\zeta}^* + \varrho \frac{\left[\sum_{i=0}^{t-1} \boldsymbol{g}(\boldsymbol{x}(i))\right]^+}{\left\|\left[\sum_{i=0}^{t-1} \boldsymbol{g}(\boldsymbol{x}(i))\right]^+\right\|}. \tag{52}$$

Recall that even if $\mathcal{D}$ is not bounded, the iterates are still assumed to be bounded (cf. Assumption 2), and in particular, to lie in the set $\mathcal{D}_{\text{ball}}$ given by (21). It is easy to verify using the triangle inequality that $\breve{\boldsymbol{\zeta}} \in \mathcal{D}_{\text{ball}}$. Noting that $\nabla_{\boldsymbol{\zeta}} L(\boldsymbol{x}(i), \boldsymbol{\zeta}(i)) = \boldsymbol{g}(\boldsymbol{x}(i))$, a straightforward manipulation yields

$$\sum_{i=0}^{t-1} (\breve{\boldsymbol{\zeta}} - \boldsymbol{\zeta}^*)^T \nabla_{\boldsymbol{\zeta}} L(\boldsymbol{x}(i), \boldsymbol{\zeta}(i)) = \varrho \left\| \left[\sum_{i=0}^{t-1} \boldsymbol{g}(\boldsymbol{x}(i))\right]^+ \right\|. \tag{53}$$

Since $\boldsymbol{g}(\boldsymbol{x})$ is convex and $[.]^+$ is a nonnegative vector, the following implications hold:

$$\boldsymbol{g}(\bar{\boldsymbol{x}}(t)) \leq \frac{1}{t}\sum_{i=0}^{t-1} \boldsymbol{g}(\boldsymbol{x}(i)) \Rightarrow [\boldsymbol{g}(\bar{\boldsymbol{x}}(t))]^+ \leq \frac{1}{t}\left[\sum_{i=0}^{t-1} \boldsymbol{g}(\boldsymbol{x}(i))\right]^+ \Rightarrow \left\|[\boldsymbol{g}(\bar{\boldsymbol{x}}(t))]^+\right\| \leq \frac{1}{t}\left\|\left[\sum_{i=0}^{t-1} \boldsymbol{g}(\boldsymbol{x}(i))\right]^+\right\|. \tag{54}$$

Consider next combining (51), (53), and (54), and taking the sup over $\mathcal{D}_{\text{ball}}$; then

$$\left\|[\boldsymbol{g}(\bar{\boldsymbol{x}}(t))]^+\right\| \leq \sup_{\boldsymbol{\zeta} \in \mathcal{D}_{\text{ball}}} \frac{1}{\varrho t} \sum_{i=0}^{t-1} (\boldsymbol{\zeta} - \boldsymbol{\zeta}^*)^T \nabla_{\boldsymbol{\zeta}} L(\boldsymbol{x}(i), \boldsymbol{\zeta}(i))$$

$$\leq \sup_{\boldsymbol{\zeta} \in \mathcal{D}_{\text{ball}}} \frac{\|\boldsymbol{\zeta}(0) - \boldsymbol{\zeta}\|^2}{2\alpha\varrho t} + \frac{1}{t}\sum_{i=0}^{t-1} \epsilon \sup_{\boldsymbol{\zeta} \in \mathcal{D}_{\text{ball}}} \|\boldsymbol{\zeta}(i) - \boldsymbol{\zeta}\| + \frac{\alpha(B_L+\epsilon)^2}{2\varrho} + \frac{\|\boldsymbol{x}(0) - \boldsymbol{x}^*\|^2}{2\alpha\varrho t} + rB_d + \frac{\alpha(B_L+r)^2}{2\varrho}$$

which readily gives (24).

(ii) By using the convexity of $f(\boldsymbol{x})$, adding and subtracting the terms $\boldsymbol{\zeta}^T(i)\boldsymbol{g}(\boldsymbol{x}(i))$ to $f(\boldsymbol{x}_i)$ below, and invoking the definition of the Lagrangian function (10) and (43), we obtain for all $t \geq 1$

$$f(\bar{\boldsymbol{x}}(t)) - f^* \leq \frac{1}{t}\sum_{i=0}^{t-1} L(\boldsymbol{x}(i), \boldsymbol{\zeta}(i)) - f^* - \frac{1}{t}\sum_{i=0}^{t-1} \boldsymbol{\zeta}^T(i)\boldsymbol{g}(\boldsymbol{x}(i)) \tag{55}$$

$$\leq \frac{\|\boldsymbol{x}(0) - \boldsymbol{x}^*\|^2}{2\alpha t} + rB_d + \frac{\alpha(B_L+r)^2}{2} - \frac{1}{t}\sum_{i=0}^{t-1} \boldsymbol{\zeta}^T(i)\boldsymbol{g}(\boldsymbol{x}(i)). \tag{56}$$

The concavity of the Lagrangian function with respect to $\boldsymbol{\zeta}$ implies that

$$\nabla_{\boldsymbol{\zeta}}^T L(\boldsymbol{x}(i), \boldsymbol{\zeta}(i))(\boldsymbol{\zeta} - \boldsymbol{\zeta}(i)) + L(\boldsymbol{x}(i), \boldsymbol{\zeta}(i)) \geq L(\boldsymbol{x}(i), \boldsymbol{\zeta}). \tag{57}$$

Using the latter in (40) with $\boldsymbol{\zeta} = \mathbf{0} \in \mathcal{D}$, and noting that $L(\boldsymbol{x}(i), \boldsymbol{\zeta}(i)) - L(\boldsymbol{x}(i), \mathbf{0}) = \boldsymbol{\zeta}^T(i)\boldsymbol{g}(\boldsymbol{x}(i))$, it follows that for all $i \geq 0$,

$$\|\boldsymbol{\zeta}(i)\|^2 \leq \|\boldsymbol{\zeta}(0)\|^2 + 2\alpha\boldsymbol{\zeta}^T(i)\boldsymbol{g}(\boldsymbol{x}(i)) + 2\alpha\epsilon\|\boldsymbol{\zeta}(t)\| + \alpha^2(B_L+\epsilon)^2. \tag{58}$$

Summing the latter for $i = 0, 1, \ldots, t-1$, dividing by $2\alpha t$, and using the boundedness assumptions, we obtain

$$-\frac{1}{t}\sum_{i=0}^{t-1} \boldsymbol{\zeta}^T(i)\boldsymbol{g}(\boldsymbol{x}(i)) \leq \frac{\|\boldsymbol{\zeta}(0)\|^2}{2\alpha t} + \epsilon B_{\boldsymbol{\zeta}} + \frac{\alpha(B_L+\epsilon)^2}{2}. \tag{59}$$

Using (59) into (56) yields (25).

(iii) It holds for all $t \geq 1$ that

$$f(\bar{\boldsymbol{x}}(t)) = L(\bar{\boldsymbol{x}}(t), \boldsymbol{\zeta}^*) - \boldsymbol{\zeta}^{*T}\boldsymbol{g}(\bar{\boldsymbol{x}}(t)) \geq f^* - \boldsymbol{\zeta}^{*T}\boldsymbol{g}(\bar{\boldsymbol{x}}(t)) \tag{60}$$

and also

$$-\boldsymbol{\zeta}^{*T}\boldsymbol{g}(\bar{\boldsymbol{x}}(t)) \geq -\boldsymbol{\zeta}^{*T}[\boldsymbol{g}(\bar{\boldsymbol{x}}(t))]^+ \geq -\|\boldsymbol{\zeta}^*\|\,\|[\boldsymbol{g}(\bar{\boldsymbol{x}}(t))]^+\|. \tag{61}$$

Using (61) into (60), (26) follows. □

Now, focus is turned to the running average in (17). It will be helpful to consider also the running average of the Lagrange multipliers,

$$\bar{\boldsymbol{\zeta}}_\beta(t) := \frac{\sum_{i=0}^{t-1}\beta^{t-1-i}\boldsymbol{\zeta}(i)}{\sum_{i=0}^{t-1}\beta^{t-1-i}} = \frac{\sum_{i=0}^{t-1}\beta^{-i}\boldsymbol{\zeta}(i)}{\sum_{i=0}^{t-1}\beta^{-i}}, \quad t = 1, 2, \ldots \tag{62}$$

The following lemma is a result about summations that appear frequently in the analysis of averaging (17), and will be used in subsequent proofs.

**Lemma 7.** *With $S_t$ as in* (27)*, it holds for all $\boldsymbol{x} \in \mathcal{X}$, $\boldsymbol{\zeta} \in \mathcal{D}$, and $t \geq 2$ that*

(i) $$\frac{1}{S_t}\sum_{i=0}^{t-1}\frac{1}{\beta^i}(\|\boldsymbol{x}(i)-\boldsymbol{x}\|^2 - \|\boldsymbol{x}(i+1)-\boldsymbol{x}\|^2) \leq \frac{\|\boldsymbol{x}(0)-\boldsymbol{x}\|^2}{S_t} + \frac{1}{S_t}\sum_{i=1}^{t-1}\left(\frac{1}{\beta^i} - \frac{1}{\beta^{i-1}}\right)\|\boldsymbol{x}(t)-\boldsymbol{x}\|^2 \tag{63}$$

(ii) $$\frac{1}{S_t}\sum_{i=0}^{t-1}\frac{1}{\beta^i}(\|\boldsymbol{\zeta}(t)-\boldsymbol{\zeta}\|^2 - \|\boldsymbol{\zeta}(t+1)-\boldsymbol{\zeta}\|^2) \leq \frac{\|\boldsymbol{\zeta}(0)-\boldsymbol{\zeta}\|^2}{S_t} + \frac{1}{S_t}\sum_{i=1}^{t-1}\left(\frac{1}{\beta^i} - \frac{1}{\beta^{i-1}}\right)\|\boldsymbol{\zeta}(t)-\boldsymbol{\zeta}\|^2 \tag{64}$$

(iii) $$\frac{1}{S_t}\sum_{i=1}^{t-1}\left(\frac{1}{\beta^i} - \frac{1}{\beta^{i-1}}\right) = (1-\beta)\frac{1-\beta^{t-1}}{1-\beta^t}. \tag{65}$$

*Proof of Lemma 7:* (i) By simply rearranging terms in the summation, it holds for all $t \geq 2$ that

$$\begin{aligned}\frac{1}{S_t}&\sum_{i=0}^{t-1}\frac{1}{\beta^i}(\|\boldsymbol{x}(i)-\boldsymbol{x}\|^2 - \|\boldsymbol{x}(i+1)-\boldsymbol{x}\|^2)\\ &= \frac{\|\boldsymbol{x}(0)-\boldsymbol{x}\|^2}{S_t} + \frac{1}{S_t}\sum_{i=1}^{t-1}\frac{1}{\beta^i}\|\boldsymbol{x}(i)-\boldsymbol{x}\|^2 - \frac{1}{S_t}\sum_{i=0}^{t-2}\frac{1}{\beta^i}\|\boldsymbol{x}(i+1)-\boldsymbol{x}\|^2 - \frac{1}{\beta^{t-1}}\|\boldsymbol{x}(t)-\boldsymbol{x}\|^2\end{aligned} \tag{66}$$

from which (63) follows easily.

(ii) This is identical to (i).

(iii) The sum is telescopic. It follows for all $t \geq 2$ that

$$\frac{1}{S_t}\sum_{i=1}^{t-1}\left(\frac{1}{\beta^i}-\frac{1}{\beta^{i-1}}\right)=\frac{\frac{1}{\beta^{t-1}}-1}{S_t}=\frac{1-\beta^{t-1}}{\sum_{i=0}^{t-1}\beta^i}=\frac{1-\beta^{t-1}}{\frac{1-\beta^t}{1-\beta}} \tag{67}$$

which is exactly (65). □

The convergence analysis is patterned after the previous one for averaging (16), generalizing it to consider the varying weights $\beta^i$. In the ensuing proofs, just the main steps for this generalization are summarized, starting with the following counterpart of Lemma 6.

**Lemma 8.** *For all* $t \geq 2$*, it holds that*

$$\frac{1}{S_t}\sum_{i=0}^{t-1}\frac{1}{\beta^i}L(\boldsymbol{x}(i),\boldsymbol{\zeta}(i))-f^* \leq \frac{\|\boldsymbol{x}(0)-\boldsymbol{x}^*\|^2}{2\alpha S_t}+\frac{B_d^2}{2\alpha}(1-\beta)\frac{1-\beta^{t-1}}{1-\beta^t}+rB_d+\frac{\alpha(B_L+r)^2}{2}. \tag{68}$$

*Proof of Lemma 8:* Multiplying (44) with $\beta^{-i}$, summing for $i = 0, 1, \ldots, t-1$, and dividing by $S_t$, it follows that

$$\begin{aligned}&\frac{1}{S_t}\sum_{i=0}^{t-1}\frac{1}{\beta^i}L(\boldsymbol{x}(i),\boldsymbol{\zeta}(i))-\frac{1}{t}\sum_{i=0}^{t-1}\frac{1}{\beta^i}L(\boldsymbol{x},\boldsymbol{\zeta}(i))\\&\quad\leq\frac{1}{2\alpha S_t}\sum_{i=0}^{t-1}\frac{1}{\beta^i}(\|\boldsymbol{x}(i)-\boldsymbol{x}\|^2-\|\boldsymbol{x}(i+1)-\boldsymbol{x}\|^2)+\frac{1}{S_t}\sum_{i=0}^{t-1}\frac{1}{\beta^i}r\|\boldsymbol{x}(i)-\boldsymbol{x}\|+\frac{\alpha(B_L+r)^2}{2}.\end{aligned} \tag{69}$$

Using the concavity of $L(\boldsymbol{x},\boldsymbol{\zeta})$ with respect to $\boldsymbol{\zeta}$ as in the proof of Lemma 6, in combination with Lemma 7 and the boundedness assumptions, (68) follows readily. □

Next, Proposition 2 is proved using Lemma 8.

*Proof of Proposition 2:* (i) Multiplying (50) by $\beta^{-i}$, summing for $i = 0, 1, \ldots, t-1$, dividing by $S_t$, and invoking Lemmas 7 and 8, it follows that for all $\boldsymbol{\zeta} \in \mathcal{D}$,

$$\begin{aligned}\frac{1}{S_t}\sum_{i=0}^{t-1}\frac{1}{\beta^i}(\boldsymbol{\zeta}-\boldsymbol{\zeta}^*)^T\nabla_{\boldsymbol{\zeta}}L(\boldsymbol{x}(i),\boldsymbol{\zeta}(i))&\leq\frac{\|\boldsymbol{\zeta}(0)-\boldsymbol{\zeta}\|^2}{2\alpha S_t}+\frac{1}{2\alpha S_t}\sum_{i=0}^{t-1}\left(\frac{1}{\beta^i}-\frac{1}{\beta^{i-1}}\right)\|\boldsymbol{\zeta}(t)-\boldsymbol{\zeta}\|^2+\frac{\alpha(B_L+\epsilon)^2}{2}\\&+\frac{1}{S_t}\sum_{i=0}^{t-1}\frac{1}{\beta^i}\epsilon\|\boldsymbol{\zeta}(i)-\boldsymbol{\zeta}\|+\frac{\|\boldsymbol{x}(0)-\boldsymbol{x}^*\|^2}{2\alpha S_t}+\frac{B_d^2}{2\alpha}(1-\beta)\frac{1-\beta^{t-1}}{1-\beta^t}+rB_d+\frac{\alpha(B_L+r)^2}{2}.\end{aligned} \tag{70}$$

Consider the following vector:

$$\breve{\boldsymbol{\zeta}} := \boldsymbol{\zeta}^*+\varrho\,\frac{\left[\frac{1}{S_t}\sum_{i=0}^{t-1}\frac{1}{\beta^i}\boldsymbol{g}(\boldsymbol{x}(i))\right]^+}{\left\|\left[\frac{1}{S_t}\sum_{i=0}^{t-1}\frac{1}{\beta^i}\boldsymbol{g}(\boldsymbol{x}(i))\right]^+\right\|}. \tag{71}$$

The rest of the proof continues as the proof of Proposition 1. In the present case, the boundedness assumptions and Lemma 7 are invoked after taking the $\sup$ of the right-hand side of (70) over $\boldsymbol{\zeta} \in \mathcal{D}_{\text{ball}}$.

(ii), (iii) Following the proof of part (ii) of Proposition 1, a bound on the sum $\frac{1}{S_t}\sum_{i=0}^{t-1}\beta^{-i}\boldsymbol{\zeta}^T(i)\boldsymbol{g}(\boldsymbol{x}(i))$ can be obtained by multiplying (58) with $\beta^{-i}$, summing for $i=0,1,\ldots,t-1$, dividing by $2\alpha S_t$, and invoking the boundedness assumptions and Lemma 7. Part (iii) is identical to that of Proposition 1. □

Finally, Corollary 1 and Lemma 4 are proved next.

*Proof of Corollary 1:* If the set $\mathcal{D}_{\text{box}}$ is used for the projection, then (51) holds for all $\boldsymbol{\zeta}\in\mathcal{D}_{\text{box}}$. But $\mathcal{D}_{\text{ball}}\subset\mathcal{D}_{\text{box}}$ and $\breve{\boldsymbol{\zeta}}\in\mathcal{D}_{\text{ball}}$ [cf. (52)], so it is still possible to take the $\sup$ over $\boldsymbol{\zeta}\in\mathcal{D}_{\text{ball}}$ in (55). The argument also holds for the right-hand side of (70). □

*Proof of Lemma 4:* Assumption 2 implies that the sequences $\{\bar{\boldsymbol{x}}(t)\}$ and $\{\bar{\boldsymbol{x}}_\beta(t)\}$ indeed have limit points. Focusing first on averaging (16), assume the claim is not true, i.e., suppose that $\text{dist}(\mathring{\boldsymbol{x}},\mathcal{X}^*)>\varepsilon$.

There will be a subsequence $\{\bar{\boldsymbol{x}}(t_k)\}$, $k\in\mathbb{N}$ so that $\bar{\boldsymbol{x}}(t_k)\to\mathring{\boldsymbol{x}}$ as $k\to\infty$. Due to the continuity of the distance function, $\text{dist}(\mathring{\boldsymbol{x}},\mathcal{X}^*)>\varepsilon$ implies that there is an integer $k'$ and an $\varepsilon'>0$ so that

$$\text{dist}(\bar{\boldsymbol{x}}(t_k),\mathcal{X})\geq\varepsilon+\varepsilon'\quad\forall k\geq k'. \tag{72}$$

Fix an integer $l$ so that $t_l>t'$, $l>k'$. Now, since $\mathcal{X}^*$ is closed and convex (cf. Assumption 1), the distance function is convex [24, Prop. 2.2.1]. It then follows from (72) that for any $k>l$,

$$\varepsilon+\varepsilon'\leq\frac{1}{t_k}\sum_{i=0}^{t_k-1}\text{dist}(\boldsymbol{x}(i),\mathcal{X}^*)=\frac{1}{t_k}\sum_{i=0}^{t_l-1}\text{dist}(\boldsymbol{x}(i),\mathcal{X}^*)+\frac{1}{t_k}\sum_{i=t_l}^{t_k-1}\text{dist}(\boldsymbol{x}(i),\mathcal{X}^*). \tag{73}$$

Using that $\text{dist}(\boldsymbol{x}(i),\mathcal{X}^*)\leq\varepsilon$ for all $i\geq t'$ into the latter, it follows that

$$\varepsilon+\varepsilon'\leq\frac{1}{t_k}\sum_{i=0}^{t_l-1}\text{dist}(\boldsymbol{x}(i),\mathcal{X}^*)+\frac{1}{t_k}\sum_{i=t_l}^{t_k-1}\varepsilon=\frac{1}{t_k}\sum_{i=0}^{t_l-1}\text{dist}(\boldsymbol{x}(i),\mathcal{X}^*)-\frac{1}{t_k}t_l\varepsilon+\varepsilon. \tag{74}$$

Taking $k\to\infty$ leads to a contradiction, $\varepsilon+\varepsilon'\leq\varepsilon$. For the averaging (17), the proof is analogous. □